\documentclass[review]{elsarticle}

\usepackage{lineno,hyperref}
\usepackage[utf8]{inputenc}
\usepackage[T1]{fontenc}
\usepackage[table,xcdraw]{xcolor}
\modulolinenumbers[5]

\usepackage{acro}
\DeclareAcronym{ARSCA-SP}{
	short = ARSCA-SP,
	long  = {\textit{Anycast Routing, Spectrum and Core Allocation with Shortest Path}}
}
\DeclareAcronym{ASMC}{
	short = ASMC,
	long  = \textit{Allocation with Strengthened Medium Core}
}
\DeclareAcronym{BPSK}{
		short = BPSK,
		long  = \textit{Binary Phase-Shift Keying}
	}
	\DeclareAcronym{CaP}{
		short = CaP,
		long  = \textit{Crosstalk-aware provisioning strategy with dedicated path protection}
	}
	\DeclareAcronym{EON}{
		short = EON,
		long  = \textit{Elastic Optical Network}
	}
	\DeclareAcronym{FA-RSSMA}{
		short = FA-RSSMA,
		long  = {\textit{Fragmentation-Aware Routing, Spectrum, Spatial Mode and Modulation Format Assignment}}
	}
	\DeclareAcronym{FA-RSSMA-CA}{
		short = FA-RSSMA,
		long  = {\textit{Fragmentation-Aware Routing, Spectrum, Spatial Mode and Modulation Format Assignment with Congestion Avoidance}}
	}
	\DeclareAcronym{FF-CASC}{
		short = FF-CASC,
		long  = \textit{First Fit Crosstalk-Aware Spectrum Compactness}
	}
	\DeclareAcronym{FF-MRC}{
		short = FF-MRC,
		long  = \textit{First Fit Multidimensional Resource Compactness}
	}
	\DeclareAcronym{FIPPMC}{
		short = FIPPMC,
		long  = \textit{Failure-Independent Path Protecting for MultiCore network}
	}
	\DeclareAcronym{MCF}{
		short = MCF,
		long  = \textit{Multi-Core Fiber}
	}
	\DeclareAcronym{MFF}{
		short = MFF,
		long  = \textit{Modulation Format Fixed}
	}
	\DeclareAcronym{MFS}{
		short = MFS,
		long  = \textit{Modulation Format Switching}
	}
	\DeclareAcronym{MIFMC}{
		short = MIFMC,
		long  = \textit{Minimum Interference and Failure-independent path protecting for MultiCore networks}
	}
	\DeclareAcronym{PLM}{
		short = PLM,
		long  = \textit{photonic-lantern multiplexer}
	}
	\DeclareAcronym{PMP}{
		short = PMP,
		long  = \textit{Phase-Matching Points}
	}
	\DeclareAcronym{QoT}{
		short = QoT,
		long  = \textit{Quality of Transmission}
	}
	\DeclareAcronym{QPSK}{
		short = QPSK,
		long  = \textit{Quadrature Phase-Shift Keying}
	}
	\DeclareAcronym{RF-CASC}{
		short = RF-CASC,
		long  = \textit{Random Fit Crosstalk-Aware Spectrum Compactness}
	}
	\DeclareAcronym{RF-MRC}{
		short = RF-MRC,
		long  = \textit{Random Fit Multidimensional Resource Compactness}
	}
	\DeclareAcronym{RMLSA}{
		short = RMLSA,
		long  = {\textit{Routing, Modulation Level and Spectrum Allocation}}
	}
	\DeclareAcronym{RMSCA}{
		short = RMSCA,
		long  = {\textit{Routing, Modulation, Spectrum and Core Allocation}}
	}
	\DeclareAcronym{ROADM}{
		short = ROADM,
		long  = \textit{Reconfigurable Add-Drop Multiplexer}
	}
	\DeclareAcronym{SCF}{
		short = SCF,
		long  = \textit{Single-Core Fiber}
	}
	\DeclareAcronym{SDM}{
		short = SDM,
		long  = \textit{Spatial Division Multiplexing}
	}
	\DeclareAcronym{SDM-EON}{
		short = SDM-EON,
		long  = \textit{Spatial-Division Multiplexing Elastic Optical Network}
	}
	\DeclareAcronym{SPSA}{
		short = SPSA,
		long  = \textit{Shortest Path with Cumulative Spectrum Availability}
	}
	\DeclareAcronym{WSS}{
		short = WSS,
		long  = \textit{Wavelength-Selective Switch}
	}
	\DeclareAcronym{XT}{
		short = XT,
		long  = \textit{Crosstalk}
	}

\journal{Journal of \LaTeX\ Templates}

\begin{document}

\begin{frontmatter}

\title{A survey on Crosstalk and Routing, Modulation Selection, Core and Spectrum Allocation in Elastic Optical Networks}

\author{Ítalo Brasileiro}
\author{Lucas Costa}
\author{André Drummond}

\begin{abstract}
Elastic Optical Networks (EON) emerge as a viable solution to supply the current growing demand for bandwidth. With the application of multi-core fibers (MCF) in EON links, it is possible to increase the availability of spectral resources. An EON network with MCF enables Space-Division Multiplexing (SDM), allowing the use of more resources in the fibers and increasing the capacity of attending circuit requests. However, the use of SDM brings some problems of interference between the circuits of a fiber, with greater emphasis on crosstalk interference. In this paper, some important concepts around EON are presented, along with the characterization of SDM technology. The Routing, Modulation, Spectrum and Core Allocation (RMSCA) problem is also characterized, and some solutions currently found in the literature are cited. After, the impact of crosstalk interference is discussed, and which elements are responsible for its occurrence. The paper is concludes with an evaluation of the state of the art, and the discrimination of the main points found from the study of papers related to the SDM-EON scenario.
\end{abstract}


\end{frontmatter}


\section{Introduction}\label{sec:intro}

Currently, efforts have been applied to develop new technologies to enable greater transmission capacity in large transport networks. In this context, \ac{EON} \cite{Jinno:09} have gained prominence, since the use of light as a data vector allows to achieve high transmission rates. In addition, the EON allows the establishment of multiple circuits in a single fiber by allocating different light frequency ranges. The EON presents the optical spectrum divided into frequency ranges of 12.5 GHz, named slots. Thus, the slots can be grouped, forming larger capacity transmission channels that allow the establishment of circuits of greater bandwidth requirement. Currently, most of the paper considers an average of 320 slots in each fiber \cite{Tode:14}, \cite{Ajmal:15}.

The optical signal can use different modulation formats by manipulating characteristics of the lightwave, such as amplitude and phase. The combination of different levels of amplitude and phase applied to the optical signal enables the transmission of a larger number of bits per symbol when compared to the traditional model of one-bit per symbol \ac{BPSK}. Therefore, the choice of the most appropriate modulation format, the choice of the route to be used and the slot range allocable to the circuit has become a problem well discussed in the literature of the EON, known as \ac{RMLSA} \cite{Drummond:17}.

The \ac{RMLSA} problem can also be modeled to consider the interference of the physical environment in transmission \cite{Beyranvand:13}. In this context, the model is closer to reality because some restrictions are added, such as the reach limitation of the modulation formats, the interference caused by the fiber type used as propagation medium, and the interference that occurs between the circuits in the same fiber.

The optical fibers considered in the traditional EONs have a single core and are referred to as \ac{SCF}. Recently, some authors have hypothesized the use of a different type of optical fiber, called \ac{MCF} \cite{Richardson:13}. The \ac{MCF} introduce a new dimension into the RMLSA problem, since multicore fibers have more than one core (usually 7 or 12), and each core has its own set of slots. Superficially, each \ac{MCF} is operated as a pool of single-core fibers.

The use of \ac{MCF}s enables the occurence of \ac{SDM}, which results in an increase of available spectral resources. Considering an \ac{EON} with \ac{MCF}, the \ac{RMLSA} problem will present another component, characterized as \textit{core choice}. Some papers refer to this new approach as \ac{RMSCA} problem \cite{Ajmal:15}.

To ensure the application cost, the use of \ac{MCF} with $n$ cores should obtain the same performance when compared to an pool of $n$ \ac{SCF}. Thus, there is reduction in the monetary cost. However, to achieve the same performance of coupled $n$ \ac{SCF}, it is necessary to reduce the interference that occurs between the \ac{MCF} cores. Among the interferences, what stands out most is the \textit{crosstalk}, and its intensity depends on the symbol rate, the modulation used and especially on the physical characteristics of the used fiber \cite{Rademacher:17}. Achieve low crosstalk and high core density is one of the main challenges for \ac{MCF} scenarios \cite{Fini:11}.

In \cite {Richardson:13} an evaluation of the evolution of transmission capacity in optical fibers is made. The authors state that the concept of \ac{SDM} is as old as the emergence of fiber-optic communication, but the current development of technologies that allow the application of \ac{SDM} has aroused interest in the scientific community. In \cite{Amaya:13} a demonstration of the first \ac{EON} with spectral spatial division is presented, with the use of an \ac{MCF} with 7 cores. The authors assemble a network of 4 nodes and 5 links (approximately 3 km each), and show the feasibility of using \ac{MCF} in optical network scenarios. The authors also present results to show the occurrence of \textit{crosstalk} and other interferences from the physical medium. In \cite{Pouria:16}, the authors investigate the performance of the different modulation formats in the \ac{SDM} scenario. They also evaluate the performance of some switching models (independent switching,  joint-switching, and fractional-joint switching) for \ac{MCF}. In \cite{Rui:16} an evaluation of different traffic aggregation policies is made in \ac{EON} scenarios with \ac{SDM}.

As shown, many papers in literature highlight the feasibility and high performance of application of \ac{MCF} in \ac{EON}. This article presents a survey of the literature around \ac{SDM-EON} found in the main publication vehicles, focusing mainly on crosstalk and \ac{RMSCA} problem. The aim is to check the state of the art and highlight research opportunities. 

This paper is organized as follows: The Section \ref{sec:sdm} presents some features of SDM technology; 
Section \ref{sec:tecnologias} presents some proposed equipment to support \ac{SDM-EON}; Section \ref{sec:crosstalk} defines some characteristics and evaluation of \textit{crosstalk} interference;
Section \ref{sec:rmsca} presents the definition of the \ac{RMSCA} problem and the proposed solutions found in the literature; finally, Section \ref{sec:conclusao} presents the challenges, conclusions and some proposals for future work.

\section{Spatial Division Multiplexing in Elastic Optical Networks} \label{sec:sdm}

This section presents some definitions around \ac{SDM} on elastic optical networks.

\begin{figure}[!ht]
	\centering
	\includegraphics[width=0.8\textwidth]{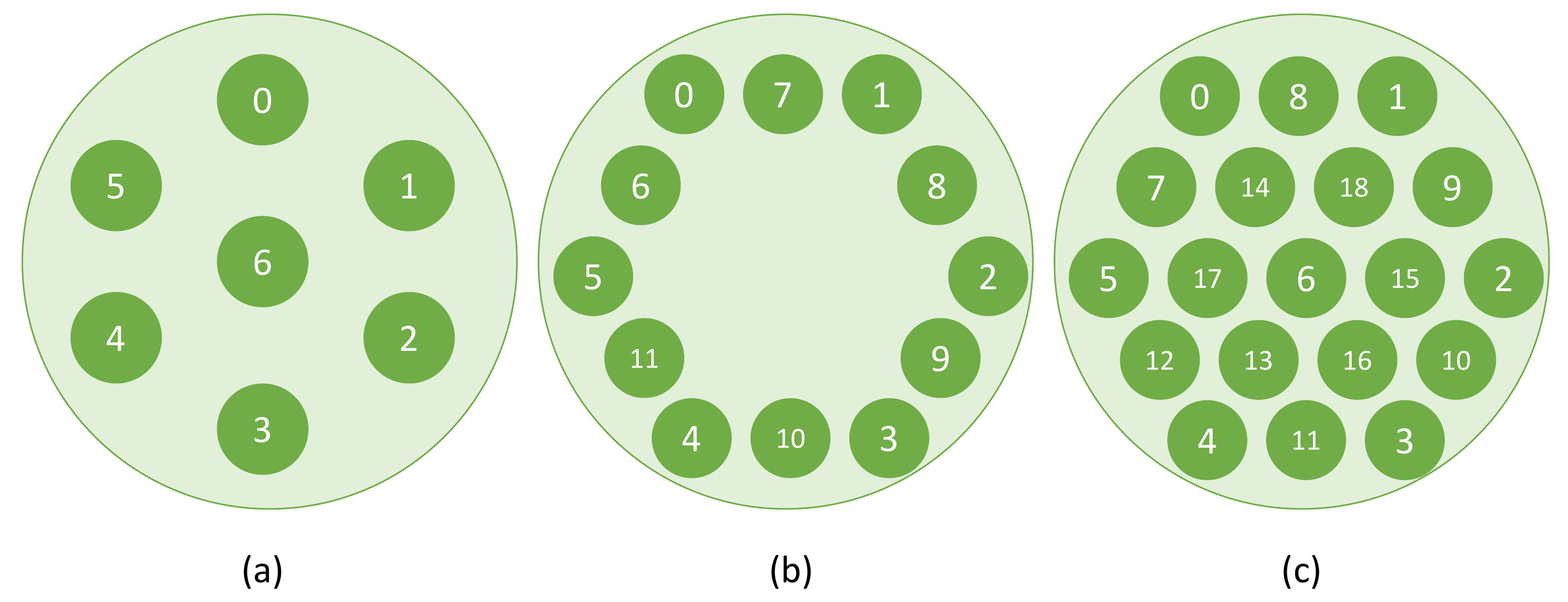}
	\caption{Multi core fiber, with (a) 7, (b) 12 and (c) 19 cores.}
	\label{fig:mcf}
\end{figure}

Currently, is observed the growth of interest in \ac{MCF} \cite{Richardson:13}. \ac{MCF}s have more cores in the fiber, unlike the traditional single-core fibers. Each core is treated as a fiber with its own set of slots. Thus, it is possible to explore additional channels in the spatial domain, which increases the transmission capacity \cite{Amaya:13}. This characteristic provided by the \ac{MCF} is called spatial-division multiplexing, and the elastic optical networks constituted by \ac{MCF} are called \ac{SDM-EON}. Figure \ref{fig:mcf} shows some examples of multi-core fibers.

The first impression is that the use of \ac{MCF}s with more cores has more advantages, due to the greater availability of resources. However, the main factor of signal interference in the \ac{MCF} is the leakage of a fraction of the signal power from a given core to its neighboring core. This phenomenon, called \textit{crosstalk} (discussed in Section \ref{sec:crosstalk}), makes it impracticable to allocate some slots, due to the great interference caused by the active circuits in the neighboring cores. Thus, in order to enable the use of \ac{MCF}s with a greater amount of cores, the development of fibers that provide smaller crosstalk between neighboring cores is required \cite{Takenaga:11-2}, \cite{Takenaga:11}.

In most of the papers found in the literature, 7-core fibers (Figure \ref{fig:mcf} (a)) are used, arranged in a hexagonal array \cite{Fujii:2014}, \cite{savva:18}. In this configuration, the central core presents 6 neighbors, and consequently suffers greater impact of \textit{crosstalk}. The peripheral cores (0, 1, 2, 3, 4 and 5 of Figure \ref{fig:mcf} (a)) have only 3 neighbors each. 12-core fibers present cores ring-like arrangement (Figure \ref{fig:mcf} (b)). In this scenario, each core has only 2 neighbors, and all cores have the same \textit{crosstalk} mean value. Fibers with 19 cores (Figure \ref{fig:mcf} (c)) have up to 6 neighbors per core, resulting in a higher incidence of \textit{crosstalk}. Still, \ac{MCF} with more cores can be used over smaller distances. For example, a \ac{MCF} with 19 cores and diameter of $200 \mu$ has high crosstalk, and fiber length limited to values close to 10 km \cite{Richardson:13}.

\begin{figure}[!ht]
	\centering
	\includegraphics[width=0.7\textwidth]{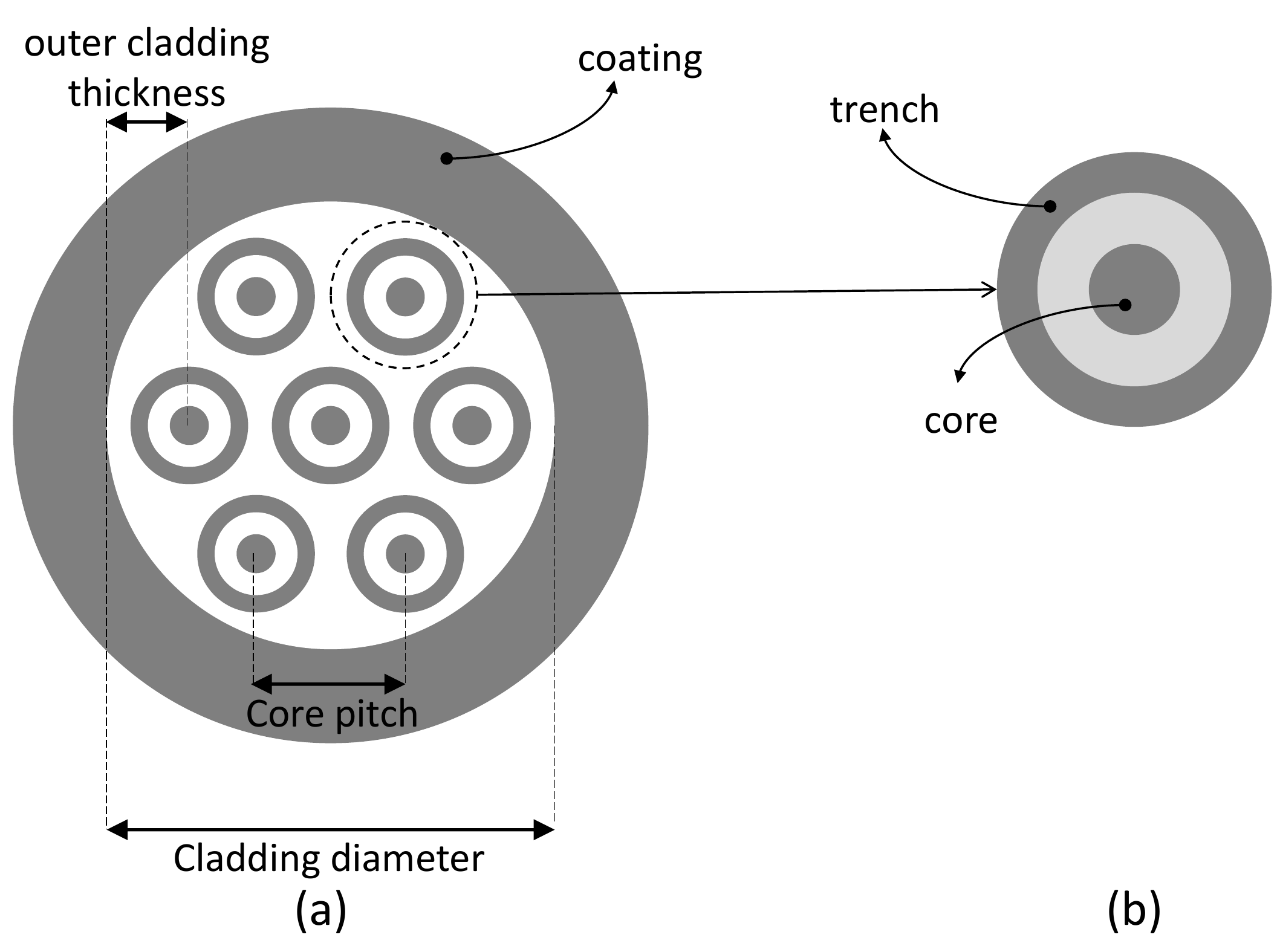}
	\caption{Layout of the elements of (a) a trench-assisted MCF and (b) of one core.}
	\label{fig:composicaoFibra}
\end{figure}

In addition to the number of cores, the arrangement of the cores and the physical characteristics of the fiber also have a strong impact on the interference between the cores. Figure \ref{fig:composicaoFibra} shows the layout of the elements of a trench-assisted MCF model.

The use of trench-assisted \ac{MCF} results in a reduction in the effects of \textit{crosstalk}. The overlap of the power of adjacent cores will be much smaller, because the trench (Figure \ref{fig:composicaoFibra}) reduces power leakage in each core. The crosstalk of a trench-assisted \ac{MCF} is around 20 dB smaller than that found in a standard \ac{MCF} \cite{Takenaga:11}. Interference between cores can be reduced by increasing the spacing between them (which reduces the number of cores, since the diameter of the fiber does not increase proportionally) or by improving the confinement of each core, as in trench-assisted fibers \cite{Fini:11}. The \textit{crosstalk} between neighboring cores has a strong dependence on the spacing between the cores (core pitch).

In order to avoid the increase of microbending loss on the outer surface of the \ac{MCF}, the increase of outer cladding thickness was proposed \cite{Jay:10}. However, fibers with coating diameter greater than 200 $\mu m$ are inappropriate for use because they are more susceptible to fractures. Thus, a less thick outer cladding is preferable, to allow for greater core scattering, higher core density, and to maintain the fiber mechanical flexibility \cite{Takenaga:11}.

Possible values for the fiber parameters found in the literature are \cite{Hayashi:12}, \cite{Hayashi:12-2}, \cite{Takenaga:11-2}: core pitch: 40.7 to 51 $\mu m$; cladding diameter: 144.6 to 188 $\mu m$; outer cladding thickness: 31.6 to 47.7 $\mu m$; coating diameter: 256 to 334 $\mu m$.

The following section highlights some proposals of equipment and technologies that make possible the application of \ac{MCF}.

\section{Support technologies to SDM-EON} \label{sec:tecnologias}

\begin{figure}[!ht]
	\centering
	\includegraphics[width=0.8\textwidth]{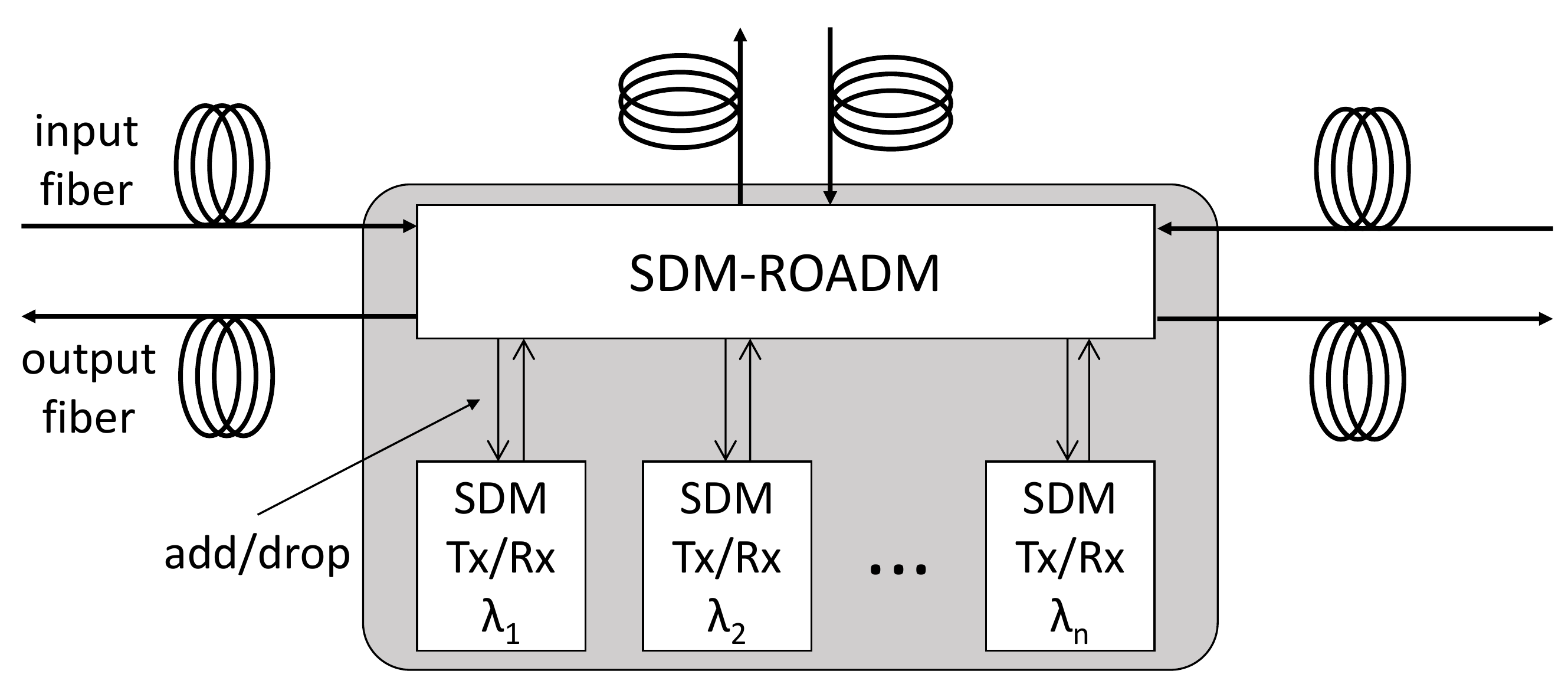}
	\caption{Potential architecture of a \ac{SDM} node, adapted from \cite{Richardson:13}.}
	\label{fig:arquiteturaROADM}
\end{figure}

Equipment that allows the circuit switching between different cores can bring significant innovation to the \ac{SDM-EON} scenario. The use of \ac{MCF}s, and consequently the expansion of the link transmission capacity, coupled with the greater flexibility of the switching between cores, leads to a relaxation of the RMSCA problem constraints. However, few papers in the literature attempt to propose a system model adapted to the scenario of \ac{SDM-EON} \cite {Richardson:13}, \cite{Marom:17}. Figure \ref{fig:arquiteturaROADM} presents a node model with support for SDM fibers.

\begin{figure}[!ht]
	\centering
	\includegraphics[width=0.85\textwidth]{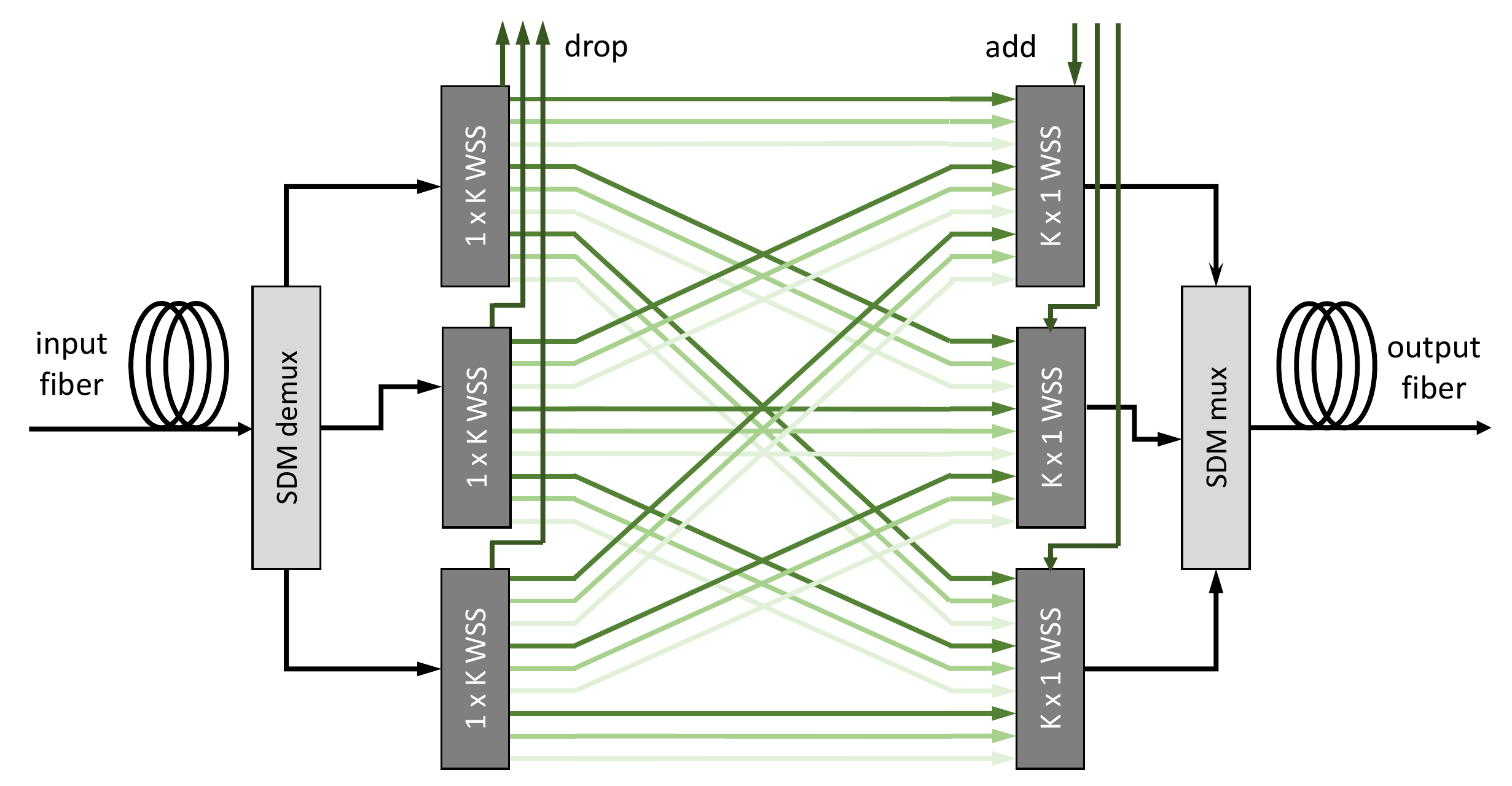}
	\caption{Potential architecture of a \ac{SDM} switch, adapted from \cite{Marom:17}.}
	\label{fig:arquiteturaSurvey}
\end{figure}

Current optical networks have flexibility due to the use of \ac{ROADM}, which allows the establishment of independent lightpaths within an optical fiber, as well as making it possible to switch them when necessary. It is considered that future \ac{SDM-EON} will have this same flexibility in routing. The Figure \ref{fig:arquiteturaROADM} presents a \ac{ROADM} adapted to \ac{SDM} scenario (SDM-ROADM), which performs the circuit switching between fiber cores, besides the add/drop function to the transmitters and receivers (Tx and Rx, respectively) \cite{Richardson:13}. Figure \ref{fig:arquiteturaSurvey} presents another switch architecture proposal that considers \ac{SDM} technology.

\begin{figure}[!ht]
	\centering
	\includegraphics[width=0.8\textwidth]{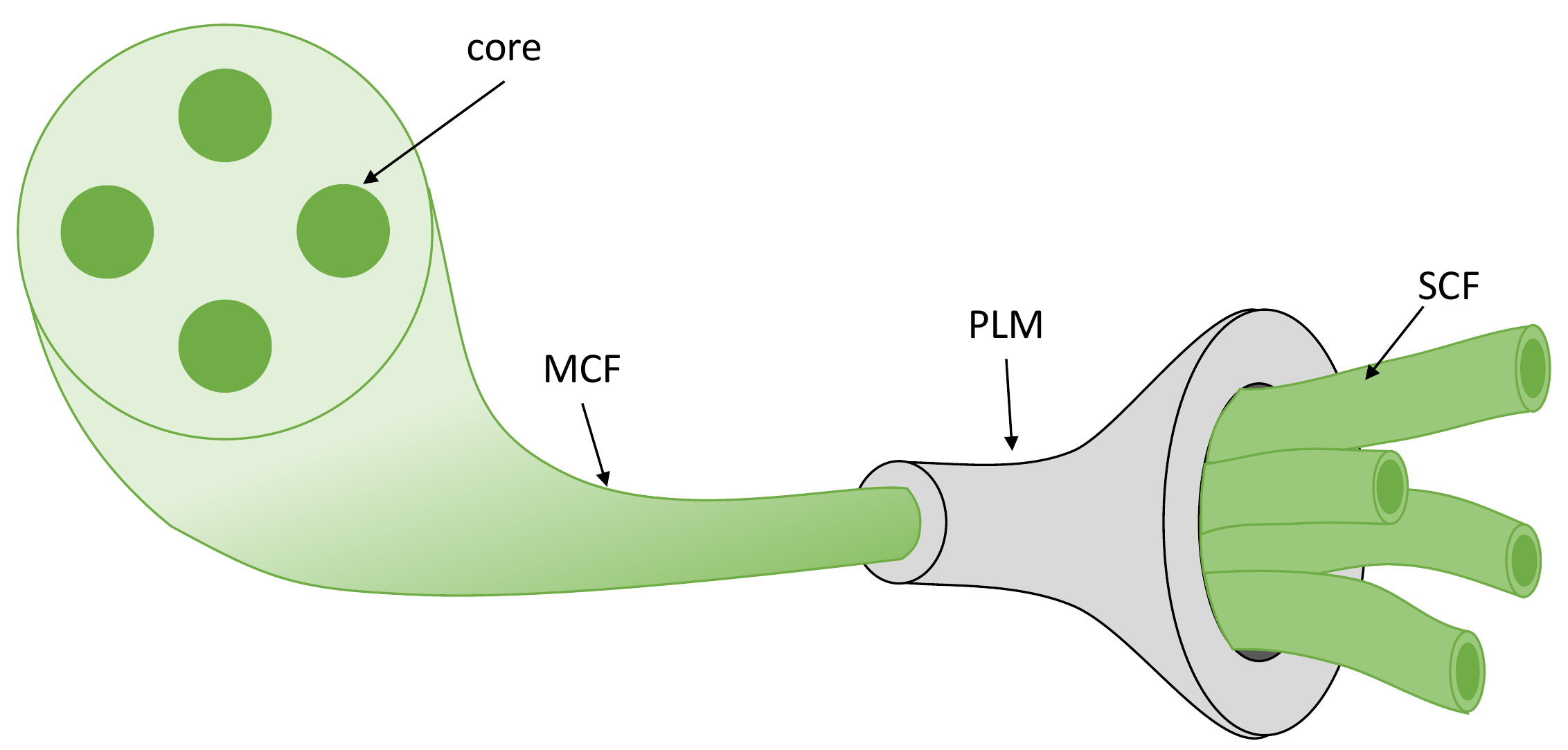}
	\caption{Photonic-lantern multiplexer \cite{Fontaine:13}.}
	\label{fig:lantern}
\end{figure}

When crossing an optical node, the input fiber passes through a spatial demultiplexer (SDM demux), which performs the separation of the spatial channels (cores). After the separation process, \ac{SCF} are used to keep each core of the input fiber, and each \ac{SCF} is directed to a \ac{WSS}. The main function of \ac{WSS} is to make the switching in a lower granularity, being able to redirect each circuit of the \ac{SCF} independently. At this point, the complexity grows with increase of the number of output ports present in the \ac{WSS}. After being switched to the appropriate port, the circuit can be directed to the current node (drop) or follow the route to another node. In this case, it is directed to a \ac{WSS}, which adds the circuit in question to the \ac{SCF} corresponding to the appropriate core (not necessarily the same core of the input fiber), which will be multiplexed, and together with the other \ac{SCF} form the output \ac{MCF} of the node. Some equipment can be adapted for the role of \ac{SDM} mux/demux, such as \ac{PLM} \cite{Fontaine:13}, which compresses multiple low capillary indexes \ac{SCF} from $n$ separate cores to a fiber with $n$ cores \cite{Fontaine:13}. The figure \ref{fig:lantern} illustrates a \ac{PLM}.

The mentioned equipment is still in development, and is not yet available for the creation of an SDM-EON. The following section describes the crosstalk interference and the equation used on the calculation, with evaluation of some crosstalk scenarios.

\section{Crosstalk} \label{sec:crosstalk}

\begin{figure}[!ht]
	\centering
	\includegraphics[width=1\textwidth]{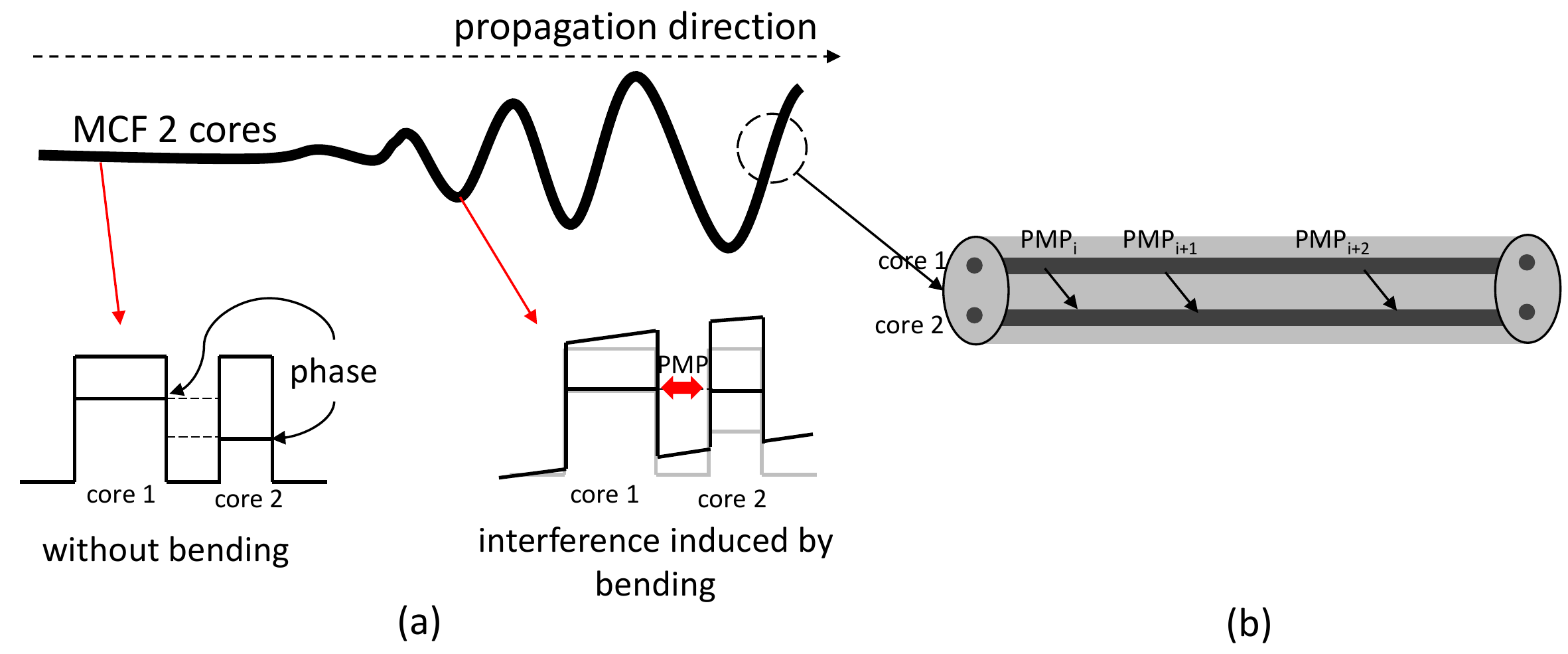}
	\caption{(a) Crosstalk occurence in a \ac{PMP}, adapted from \cite{Fini:11} and (b) different \ac{PMP}s along the fiber.}
	\label{fig:pmp}
\end{figure}

The \textit{crosstalk} is seen as the main interference on \ac{MCF}. It occurs mainly at discrete points along the fiber, called \ac{PMP}. The force of interaction between two cores occurs even with small perturbations in the fiber (radius of curvature $> 1m$) \cite{Fini:11}. Figure \ref{fig:pmp} shows an example of PMP occurrence in a fiber and (b) power loss in several fiber \ac{PMP}s \cite{Fini:11}, \cite{Rademacher:17}.

The crosstalk (after fiber propagation and installation) is a statistical value, since the occurrence of \textit{crosstalk} in the \ac{PMP}s is influenced by the phase-shift variations between the neighboring cores, and because the phase displacement is easily varied by small changes in the conditions of the fiber, such as curvature and torsion \cite{Hayashi:12}.

Observing Figure \ref{fig:mcf} (a), a circuit allocated to core 0, slots 2, 3 and 4 would suffer \textit{crosstalk} interference if circuits were allocated in cores 1, 5 or 6, in slots 2, 3 and 4. The signal of a circuit becomes noise if its \textit{crosstalk} level exceeds the threshold allowed by the network. The equation \ref{eq:crosstalk} shows how the \ac{XT} \cite{Ruijie:16-2} is calculated.

\begin{eqnarray} \label{eq:increase} 
h = \frac{2k^{2}r}
{\beta w_{tr}},
\end{eqnarray}

\begin{eqnarray} \label{eq:crosstalk} 
XT = \frac{n - n  .  exp[-(n + 1) . 2hL]}
{1 + n . exp[-(n + 1) . 2hL]},
\end{eqnarray}

\begin{figure}[!ht]
	\centering
	\includegraphics[width=0.85\textwidth]{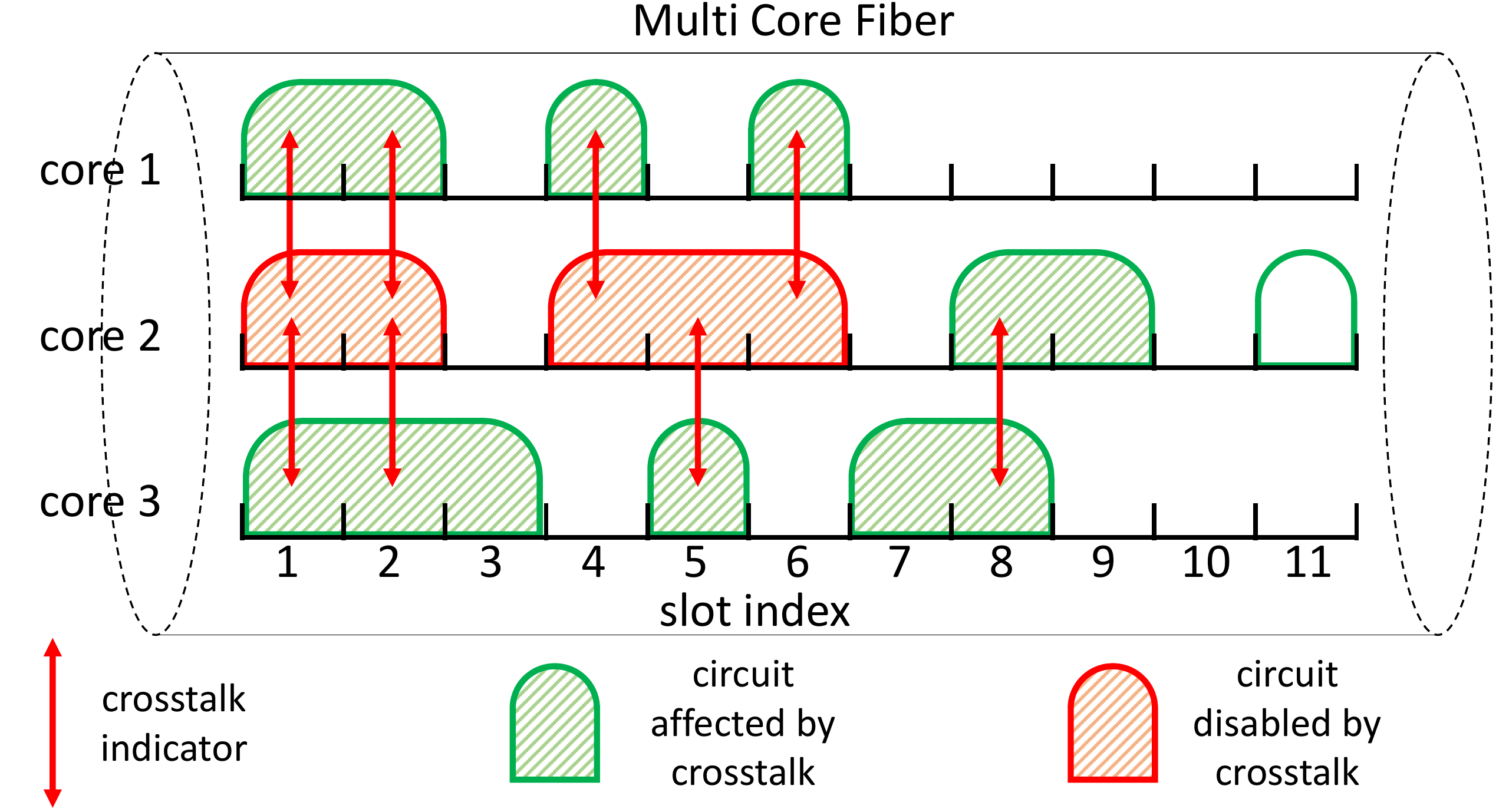}
	\caption{Crosstalk occurence in 3-core fiber.}
	\label{fig:crosstalk}
\end{figure}

In equation \ref{eq:increase}, $h$ is the increment of \textit{crosstalk} per unit length, $k$ is the fiber coupling coefficient, $r$ is the curvature radius of the fiber, $\beta$ is the propagation constant and $w_{tr}$ is the distance between cores (core pitch). In Equation \ref{eq:crosstalk}, $n$ is the number of adjacent cores (neighboring cores) and $L$ is the fiber length. The Figure \ref{fig:crosstalk} presents a demonstration of the occurrence of \textit{crosstalk} on a 3-core fiber \cite{Yuanlong:16}.

It is important to note that the occurrence of \textit{crosstalk} is more intense between adjacent cores. In figure \ref{fig:crosstalk} it is observed that the core 2 suffers greater crosstalk interference, since the two adjacent cores (1 and 3) present some circuits allocated in similar intervals of slots, as for example the slots 1, 2, 4, 5 and 6. Therefore, in the \ac{SDM} scenario, the circuit allocation should consider the index of allocated slots in the neighboring cores, in order to avoid \textit{crosstalk}. This intensifies the problem of spectral fragmentation, since certain slots can not be allocated, in order to avoid interference.

Crosstalk levels below -25 dB are required to avoid significant penalties for transmission \cite{Richardson:13}. Circuits that reach a \textit{crosstalk} level above the threshold present problems in signal interpretation on the destination node. Therefore, it is indicated that the circuits allocation does not occur in slots whose index is the same of slots allocated in neighboring cores, avoiding interference. This form of spectral allocation results in greater disorganization of the allocable spectrum, which contributes to the fragmentation of the spectrum.

It is also possible to consider the other physical layer interference effects in addition to \textit{crosstalk}. In \cite{Roberto:15}, the authors call \textit{3D} the \ac{EON} that use the three domains: temporal, spectral and spatial. The authors propose two physical impairment-aware algorithms (\ac{FA-RSSMA} and \ac{FA-RSSMA-CA}), and evaluate performance compared to SP-FF (\textit{Shortest Path} and \textit{First Fit}). The \ac{QoT} of the signal is also considered.

The paper presented in \cite{Ajmal:15} defines independent crosstalk thresholds for each modulation format, using an empirical model proposed in \cite{Politi:12}. However, when used along with the distance reach of the modulations found in the literature \cite{Drummond:17}, it is noted that the distance threshold is overestimated when compared to the crosstalk threshold. As a result, experiments were carried out to verify the mean crosstalk value for the distance thresholds found in the literature for the modulation formats.

Simulations were performed with the ONS simulator \cite{ONS-SBRC}. The independent replication method was employed to generate confidence intervals with 95\% confidence level. Each simulation run involved 100.000 requests with the following connection requests rates: 10, 20, 40, 80, 160 e 200 Gbps, all with the same arrival probability. A load point of $1,000$ Erlangs was evaluated, and 5 replications were performed. Connection requests follow a Poisson process with the mean holding time of 600 seconds, according to a negative exponential distribution and uniformly-distributed among all nodes-pairs. To do the crosstalk evaluation, the following values are used in Equation \ref{eq:crosstalk}: $k=4 * 10^{-4}$, $r=50mm$, $\beta = 4 * 10^{6}$ e $w_{tr} = 40 \mu m$.

A pair of nodes with one bidirectional link is used, and the link length varies with step of 1,000 from 1,000 km up to 10,000 km. The distances 250 km and 500 km were also used, once they represent the reach of 64QAM and 32QAM, respectively. The granularity of frequency slot is 12.5 GHz. The fiber is a 7-MCF (fig. \ref{fig:mcf}(a)), with 320 slots in each core. The guard band between two adjacent lightpaths is assumed to be 1 slot.

\begin{figure}[!ht]
	\centering
	\includegraphics[width=1\textwidth]{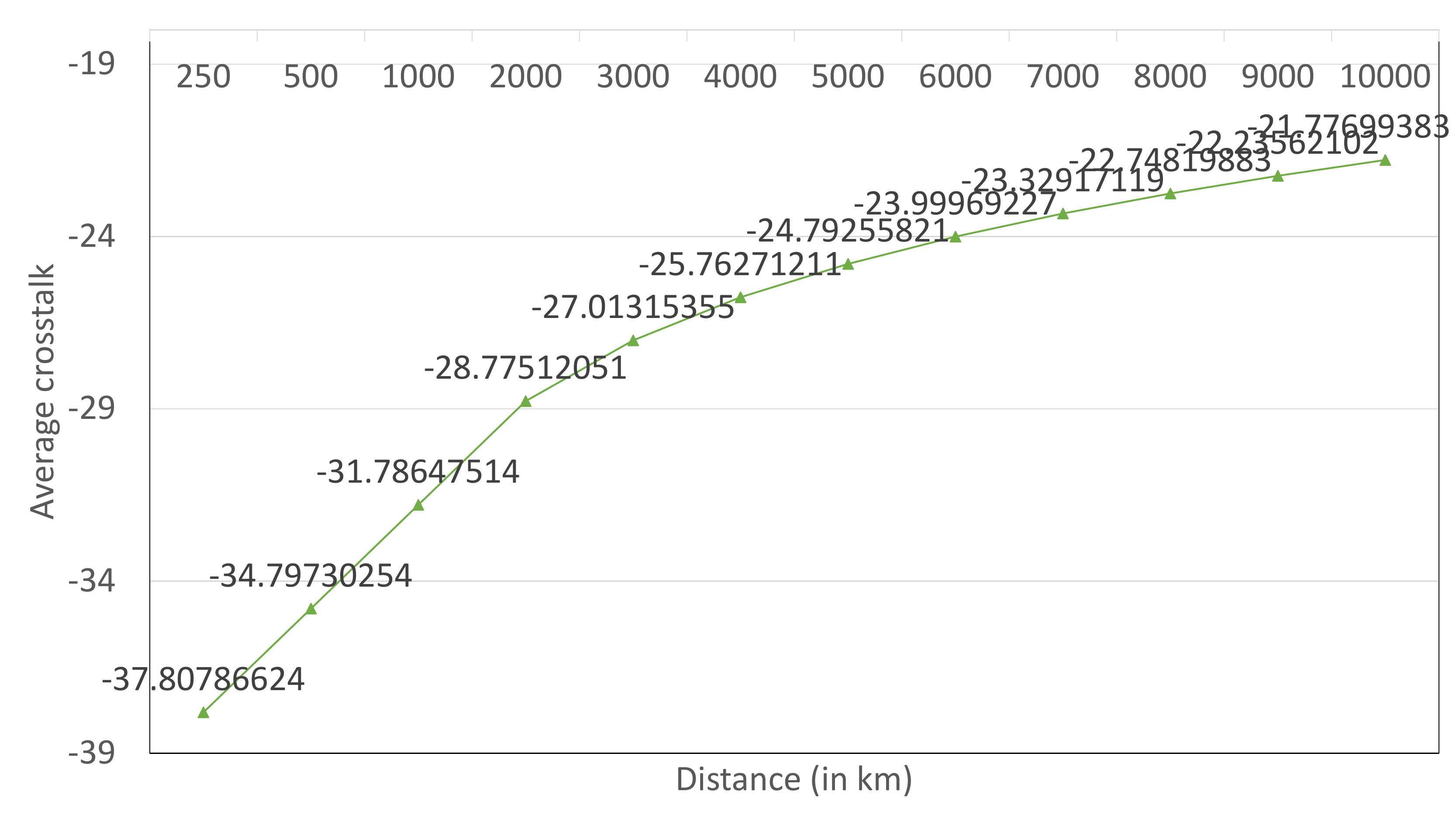}
	\caption{Crosstalk variation with increasing of distance.}
	\label{fig:XTdistancia}
\end{figure}

With the values found in the scenario evaluation of Figure \ref{fig:XTdistancia}, crosstalk thresholds were defined for the modulation formats according to their respective distance threshold. Table \ref{tab:limiarXT} presents the crosstalk thresholds proportional to the reach of the modulation formats.

\begin{table}[h!]
	\begin{center}
		\caption{Definition of modulation threshold for the respective distance reach.}
		\label{tab:limiarXT}
		\begin{tabular}{|c|c|c|c|}
			\textbf{Modulation} & \textbf{Transmission} & \textbf{Distance} & \textbf{Crosstalk}\\
			& \textbf{Capacity} 	& \textbf{Reach} 	& \textbf{Threshold}\\
			\hline
			BPSK	& 12.5 Gbps	& 8000	& -22.75\\
			QPSK	& 25 Gbps	& 4000	& -25.76\\
			8QAM	& 37.5 Gbps	& 2000	& -28.77\\
			16QAM	& 50 Gbps	& 1000	& -31.79\\
			32QAM	& 62.5 Gbps	& 500	& -34.80\\
			64QAM	& 75 Gbps	& 250	& -37.81\\
		\end{tabular}
	\end{center}
\end{table}

For calculating crosstalk, in addition to the constants derived from the physical characteristics of the MCF (such as curvature radius and core pitch), the two variables that must be taken into account at the time of allocation are the distance $L$, which is obtained by the chosen route length, and the number of neighbors $n$ to the core chosen for allocation. When the number of neighbors is taken into account, the papers found in the literature are divided into two groups: the first group considers a value of fixed $n$, being $n = 3$ for the peripheral cores and $n = 6$ for the central core \cite{Ajmal:15}, and the second group considers a dynamic $n$ value, in which are considered only the neighbors that have active circuits in the same slot index of the circuit to be allocated. \cite{Tode:16}. In this case, Figure \ref{fig:crosstalk} can be cited as an example, in which the circuit allocated in slots 8 and 9 of core 2 has $ n = 1 $, since it has a single active neighbor (slots 7 and 8 of core 3).

In addition to the problem of using a static or dynamic $ n $ value, we also highlight another point in relation to the crosstalk effect. In some papers, when establishing a new circuit, the viability of the crosstalk is also verified in circuits that are already established in neighboring cores \cite{Ruijie:16-2}. This evaluation is only done in scenarios in which the value of $ n $ is dynamic. In static $ n $ scenarios, when calculating crosstalk, the maximum neighbors capacity of the circuit (3 or 6 in the case of the central core) is already taken into account.

\begin{figure}[!ht]
	\centering
	\includegraphics[width=0.9\textwidth]{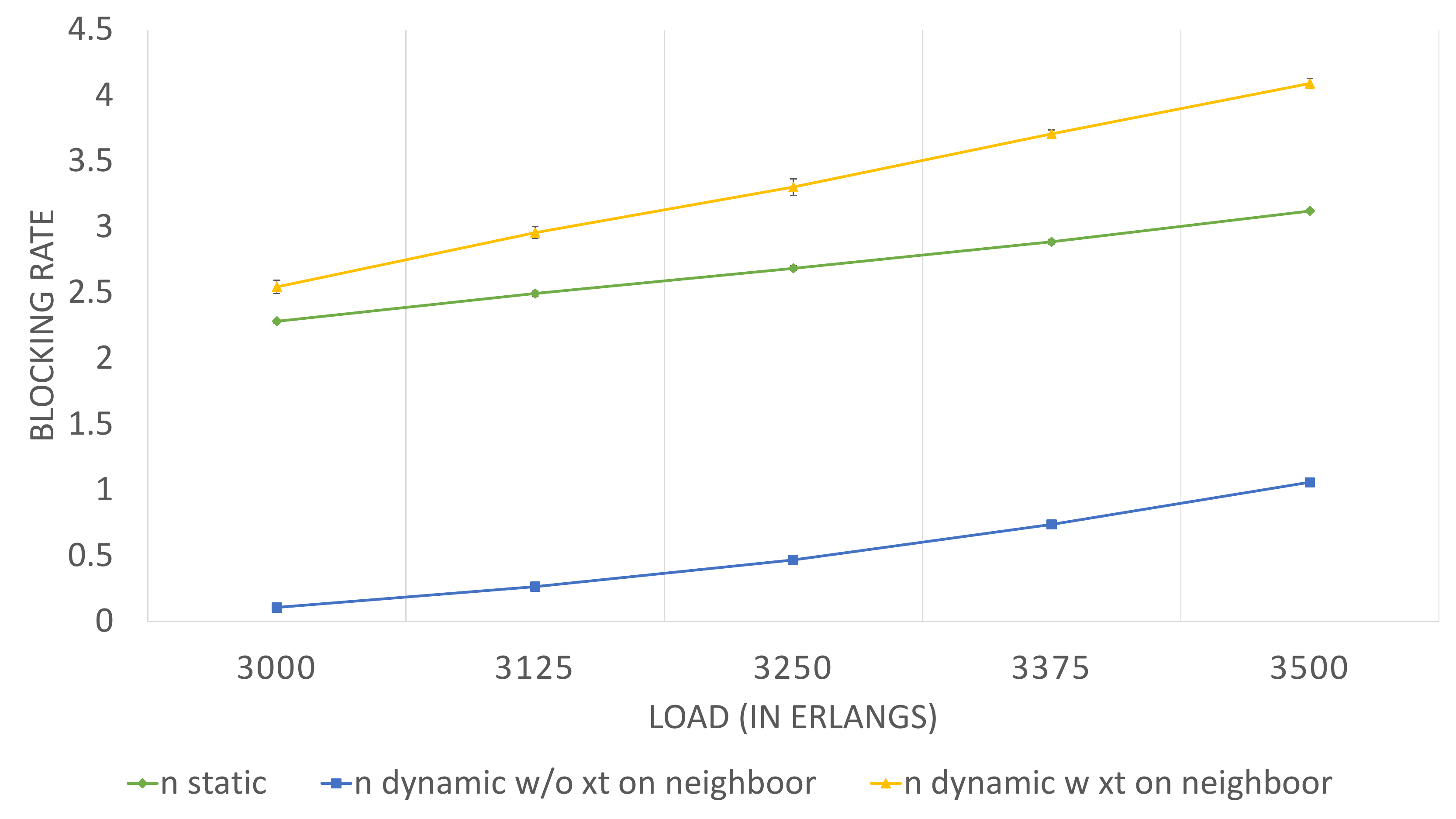}
	\caption{Blocking variation with static and dynamic values of n.}
	\label{fig:avaliacaoXT}
\end{figure}

From the modulation thresholds presented in Table \ref{tab:limiarXT}, experiments were carried out to verify the impact of the use of static and dynamic $ n $. In the case of dynamic $ n $, cases with and without re-evaluation of crosstalk were considered for the circuits already established in the network. The simulation scenario below presents the same parameters of the scenario considered for the evaluation of Figure \ref{fig:XTdistancia}. The USA topology (24 nodes and 43 3 links, detailed in Figure \ref{fig:topologias}) was used. The allocation of resources is done by policy \textit{FirstFit}, both for the choice of core and for the choice of slots. The figure \ref{fig:avaliacaoXT} shows the graphical result of the executed simulations.

For the performance evaluation shown in Figure \ref{fig:avaliacaoXT}, it is noticed that the scenario with dynamic $ n $ and without the crosstalk verification of the neighbors presents lower blocking rate. In this scenario, there is a large occurrence of cases with $ n = 0 $, which results in an extremely low crosstalk value, and allows the establishment of most of the circuits with more efficient modulations. The worst performance occurs for the scenario with dynamic $ n $ and crosstalk verification of the neighbors. As in the previous scenario, the dynamic $ n $ value allows the occurrence of cases where $ n = 0 $. When the circuits in this case are established, more efficient modulation formats are applied, since the crosstalk value is much smaller than the thresholds of the table \ref{tab:limiarXT}. Then, when the attempt to allocate resources in neighboring cores occurs, the circuit that was established previously prevents the establishment of the new circuit, since the modification of the value of $ n $ (from 0 to 1, for example), can lead the crosstalk of the circuit already established to values greater than the threshold supported by the applied modulation format.

It should be noted that the choice of the crosstalk calculation model plays an important role in the modeling of the SDM-EON scenario, since there is a large variation between the results of the models considered in the scenario evaluated. The scenario with dynamic $ n $ and no crosstalk evaluation in neighbors, there is a blocking reduction of $ 66.09\% $ when compared to the static $ n $ scenario and $ 74.13\% $ when compared to the dynamic $ n $ scenario and evaluation crosstalk in the neighbors.

The following section presents some definitions to characterize the \ac{RMSCA} problem, and comparison between some RMSCA solutions found in literature.

\section{RMSCA Problem}\label{sec:rmsca}

The establishment of circuits in optical networks requires allocation of resources, which are reserved for data transmission. In a dynamic traffic scenario, when a circuit establishment request arises, the source and destination pair $pair(s, d)$ of the new circuit is informed, in addition to the data rate to be transmitted. In the static traffic scenario, in addition to this information, the traffic matrix of the circuits to be established is also provided.

After obtaining the $pair(s, d)$, the next step is to find the appropriate route for the circuit establishment. The route represents the set of fiber links and optical nodes through which the circuit will be transmitted, in order to arrive at its destination. In order to efficiently accommodate the new circuit and save resources for future circuits, some papers choose to allocate shortest path \cite{Muhammad:14} or k-shortest paths \cite{Pouria:16} routes.

With the chosen route, the distance to be traveled by the circuit becomes known. This information is important for solving the next step in establishing the circuit: the choice of modulation \cite{Seitaro:17}. The modulation format represents the density of the optical circuit. More complex modulation formats allow the transmission of more bits per signal, while the more robust formats transmit fewer bits per signal. Thus, more complex modulation formats use fewer spectral resources, since they are able to transmit more information when compared to the more robust signals. \cite{Seitaro:17}.

The choice of the modulation format allows to define the transmission capacity of the new optical circuit, taking into account the required data rate. With this information, it is possible now to define the bandwidth that should be allocated to the circuit. In elastic optical networks, the optical spectrum of the links are arranged in small frequency slots, which are grouped together to form a transmission channel capable of containing the new circuit. Thus, the next step in establishing the circuit is the allocation of the appropriate slot interval.

\begin{figure}[!ht]
	\centering
	\includegraphics[width=0.8\textwidth]{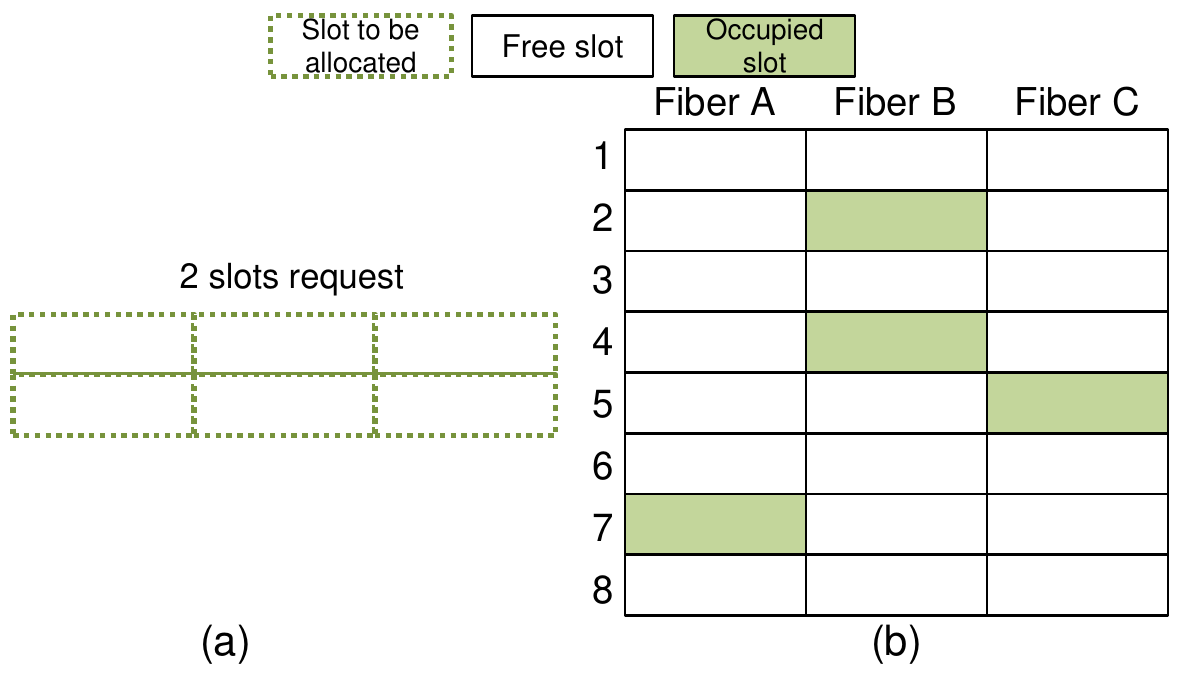}
	\caption{Continuity and contiguity restrictions blocking the establishment of a 2-slot circuit.}
	\label{fig:restricoes}
\end{figure}

It is important to emphasize that slot allocation must attend some restrictions from the optical medium. During propagation of the signal, it is preferable to keep the data transmission in the optical medium, avoiding conversion to the electronic medium, in order to reduce the use of resources and the transmission time. Therefore, it is necessary to fulfill some restrictions from the optical medium, called \textit{continuity} and \textit{contiguity} restriction. In the continuity restriction, the permanence of the optical signal in the same spectral range between the source and destination nodes becomes mandatory. Thus, when allocating a set of slots, it must be free in all links of the selected route. In contiguous restriction, it is necessary to allocate a set of slots which are adjacent to each other. With this, only one transmitter is used for each circuit, since only one spectral range is occupied. The figure \ref{fig:restricoes} illustrates a scenario in which the constraints block the establishment of a 2-slot circuit, considering the route formed by the fibers A, B and C.

The figure \ref{fig:restricoes} (a) presents a 2-slots circuit request, which must be satisfied using the resources of Figure \ref{fig:restricoes} (b). The route of the circuit was chosen in a previous stage, and it must travel through the fibers A, B and C. Considering the restrictions, it is not possible to establish the circuit: there is no set of two adjacent free slots (restriction of contiguity) maintaining the index in all three links of the route (restriction of continuity).

As a consequence of the mentioned restrictions, the presence of small free slot intervals interferes with the operation of the network, since some requests will not be answered even if there is enough free slots. This is because these slots will be scattered in the optical spectrum (as in the example of Figure \ref{fig:restricoes}), unable to be allocated due to the continuity and contiguity constraints. This problem is well discussed in the literature of elastic optical networks, and is characterized as \textit{fragmentation problem} \cite{Wang:14}.

\begin{figure}[!ht]
	\centering
	\includegraphics[width=0.8\textwidth]{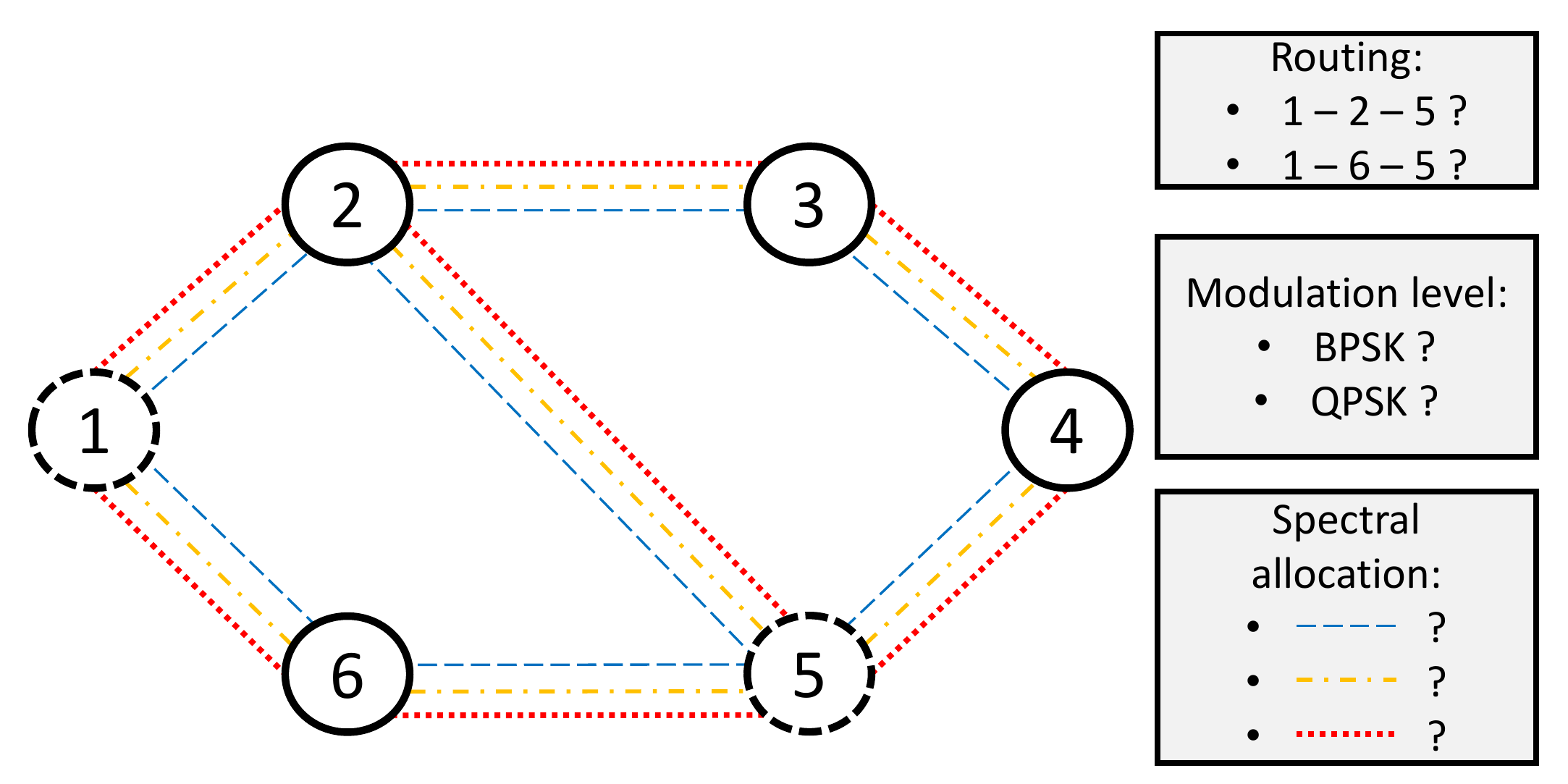}
	\caption{RMLSA problem to circuit between nodes 1 and 5.}
	\label{fig:rmlsa}
\end{figure}

The route choice, the definition of modulation format and spectral allocation are prominent problems in the literature of \ac{EON}. Together, they form the \ac{RMLSA} problem. Figure \ref{fig:rmlsa} demonstrates the \ac{RMLSA} problem in a simple network.

As shown in Figure \ref{fig:rmlsa}, the first step of the \ac{RMLSA} problem is the choice of route to be used. After solving the routing problem, the total distance to be traveled by the circuit becomes known. With this information, in the second stage the modulation format used is chosen. The choice of the modulation format (\ac{BPSK} or \ac{QPSK} in the example of Figure \ref{fig:rmlsa}) is done as a function of the route length, since more complex modulation formats symbol has a smaller range due to the greater fragility of the signal as it traverses the transmission medium. The choice of modulation format allows to decide the number of slots that will be used by the circuit. Then, in the third step, the problem of allocating the range of slots within the optical spectrum of the chosen route is solved.

With the use of \ac{MCF}, the \ac{RMLSA} problem will also cover the choice of the most suitable core for the circuit. Thus, the problem is called \ac{RMSCA} \cite{Ajmal:15}. For the core allocation phase, it is important to observe the indexes of the slots already allocated in the adjacent cores (or neighbors) to the chosen core, since the interference between cores (\textit{crosstalk}, detailed in section \ref{sec:crosstalk}) is an important factor and should be considered in studies for closer proximity to real scenarios of SDM-EON.

\begin{figure}[!ht]
	\centering
	\includegraphics[width=0.7\textwidth]{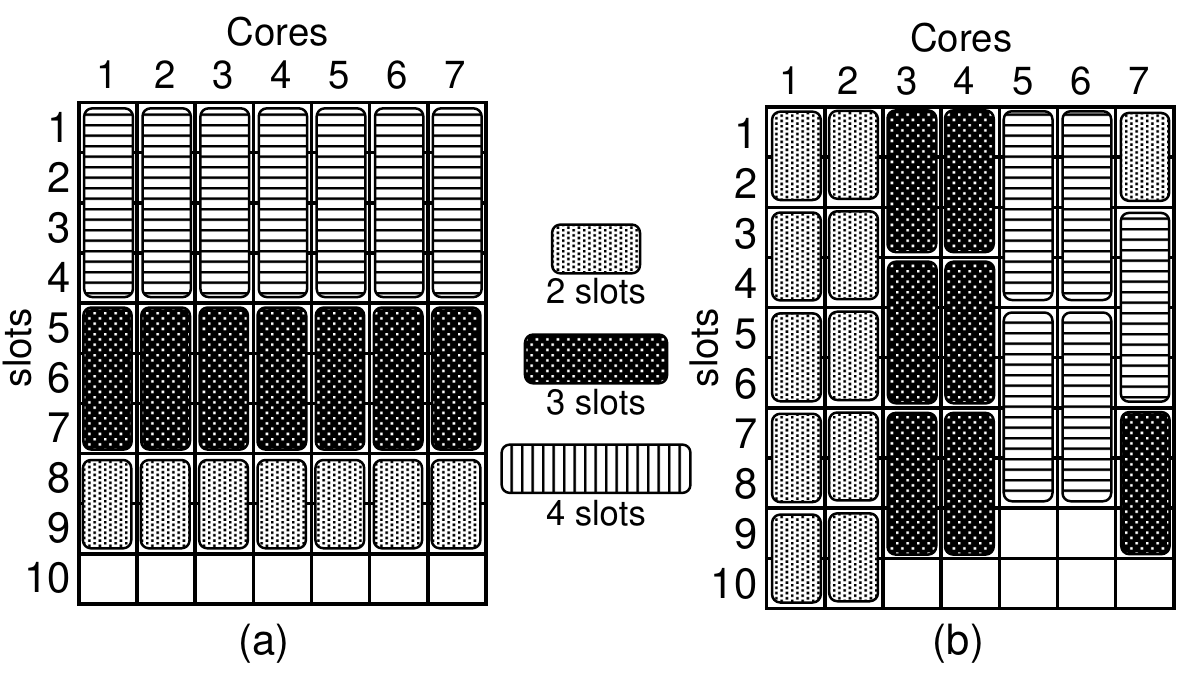}
	\caption{Circuits allocated and organizated in (a) priority by slot index and (b) priority by core.}
	\label{fig:organizacaoNucleos}
\end{figure}

To reduce the impact of the fragmentation problem on the fiber cores, some solutions proposed to the \ac{RMSCA} problem create allocation priorities \cite{Tode:16}, \cite{Fujii:2014}. The Figure \ref{fig:organizacaoNucleos} presents some allocation models with (a) priorities by slots index and (b) priorities per core.

The figure \ref {fig:organizacaoNucleos} (a) presents an example of allocations with priorities defined by slot index \cite{Tode:16}. Thus, there will be slot ranges in the spectrum which are exclusive for the allocation of specific request bandwidth. For example, in Figure \ref{fig:organizacaoNucleos} (a), slots 1 to 4, in all cores, are exclusive for 4-slots requests. Also, is defined that a range of slots, generally called \textit{common area}, will allocate circuits that can not be allocated in their respective priority area, due to unavailability of resources or fragmentation.. In the Figure \ref{fig:organizacaoNucleos} (b), the circuits are allocated primarily in specific cores \cite{Fujii:2014}. For example, 4-slot requests will be allocated primarily in cores 5 or 6. There is also a core used as \textit{common area}.

Some authors evaluate the use of \ac{MCF} in static traffic scenarios, in which the requests have origin, destination and bandwidth defined, and the traffic matrix for the topology in question. In \cite{Muhammad:14}, the authors add the \textit{crosstalk} information as constraint to the circuit establishment, and propose the algorithm \ac{SPSA} for routing, core and slot allocation. In the evaluation, it is considered a 3-core fiber. The authors found that the effects of crosstalk are attenuated with the use of fibers with greater slots availability, in which the circuits are scattered in the spectrum and the interference is reduced. In \cite{Ajmal:15}, an objective function is proposed for the choice of route, slots and cores to be allocated. The chosen resources (route, slots and core) are those that meet the \textit{crosstalk} threshold and maximize the objective function. The results show the performance of the proposed objective function, considering two different forms of modulation format selection (\ac{MFF} and \ac{MFS}).

In \cite{Zhang:16}, the algorithm \ac{ARSCA-SP} is proposed, which allocates slots closer to the beginning of the optical spectrum (\textit{First Fit}) in all cores of the network links. The algorithm looks for smaller paths and allocates slots in the core with the lowest available slot index and that respects the \textit{crosstalk} threshold. The performance comparison is done with an ILP strategy.

Some authors propose solutions for the routing, modulation, core and spectral range choice, in scenarios of \ac{EON} with \ac{MCF} and dynamic traffic, in which the requests arrival time and duration time are unknown. In \cite{Tode:13}, a SCA (Spectrum and Core Allocation) method is proposed for core and slot selection. The algorithm reserves cores for requests with a certain number of slots, and creates priority levels between the cores. A performance evaluation is made comparing the proposal with the algorithms \textit{First Fit} and \textit{Random}, in 7-core fibers. The proposal obtains a lower blocking probability, smaller crosstalk average and less fragmentation. In \cite{Tode:14}, a solution is proposed for core and slot allocation. The algorithm is compared with \textit{First Fit} and \textit{Random Fit} allocation algorithms in a 7-core fibers, and presents better performance. The proposal also has a smaller \textit{crosstalk per slot}. The comparison with \textit {First Fit} and \textit{Random} algorithms is also done in \cite{Fujii:2014}, which proposes a method for core classification and prioritization, in which cores are exclusives to a given bandwidth request. The authors use 7, 12 and 19-core fibers, and check \textit{crosstalk} through a crosstalk-by-slot (CpS) indicator, presented in Equation \ref{eq:crosstalkPerSlot} and also used in other papers \cite{Tode:14}, \cite{Oliveira:17}:

\begin{eqnarray} \label{eq:crosstalkPerSlot} 
CpS = \frac{ n_{C}}
{n_{T}},
\end{eqnarray}

where $n_{T}$ represents the number of slots occupied in the link and $n_{C}$ represents the number of occupied slots that are also occupied at the same index in adjacent cores.

In \cite{Tode:16} the \textit{Intra Area FF Assignment} algorithm is proposed, for spectrum and core allocation. The algorithm creates ``exclusive areas'' in the optical spectrum for certain bandwidths and ``common areas'' for allocation if the exclusive areas are unavailable. The algorithm uses \textit{First Fit} to allocate circuits with even number of slots and \textit{Last Fit} to allocate circuits with odd number of slots in common area. The proposal is compared to \textit{Random} and \textit{First Fit}. In \cite{Hashino:17} the concept of \textit{XT-prohibited slot} is defined, which are the free slots that can not be allocated, since they allow the increase of crosstalk to the unwanted levels.

In \cite{Oliveira:16}, the algorithm \ac{FIPPMC} is proposed for the survivability scenario in \ac{EON}. The algorithm \ac{FIPPMC} creates a list of ``candidate paths'', which corresponds to all possible combinations of route and spectrum available to the circuit. Each candidate path receives an evaluation value, which takes into account the occurrence of \textit{crosstalk}. The candidate path chosen as the primary path is the one with the lowest evaluation value, and the path for dedicated protection is the one with lowest evaluation value and links disjoint from the main path. In \cite {Oliveira:17-2} the proposal of \cite{Oliveira:16} is adapted to use shared routes for protection. The same authors, in \cite{Oliveira:17}, propose the algorithm \ac{MIFMC}, also for protection. In the proposed algorithm, a circuit is only established if there is a disjoint route to it. If the disjoint route does not exist, a new primary route is searched, and the disjoint routes are evaluated again. This procedure is repeated until all primary routes are evaluated. Already in \cite{Yuanlong:16}, the algorithm \ac{CaP} is proposed for primary route and backup route choice. The algorithm chooses two disjoint routes in the first core and available slots interval that attend the crosstalk threshold.

Some papers take into account information about the network state during the operational phase, such the optical spectrum fragmentation. A spectral fragmentation analysis is done in \cite{Ruijie:16}. The authors propose two fragmentation-aware algorithms for core and spectrum allocation: \ac{FF-MRC} and \ac{RF-MRC}. The authors compare the results with a implementation of \textit{Dijkstra} for routing and \textit{First Fit} for core and spectrum allocation. In \cite{Ruijie:16-2} the same authors add \textit{crosstalk} information to the core and spectrum allocation, and propose the \ac{FF-CASC} and \ac{RF-CASC} algorithms. In \cite{Seitaro:17} are created dedicated areas for the different request bandwidths. In addition, slot and core allocation is also fragmentation-aware.

In order to reduce network fragmentation, it is also possible to perform the spectral defragmentation procedure. In this scenario, the circuits already allocated in the network are repositioned, in order to reduce the fragments of free slots in the network links and to enable the formation of new circuits. In \cite{Meloni:16} and \cite{Imran:15} discuss the push-pull mechanism for defragmentation, in which circuits are reallocated to different indices and cores without the need for circuit shutdown. This is due to the ``slip'' of the circuit on empty slots. In \cite{Zhao:18} a defragmentation model is proposed that takes into account the SC (spectrum compactness) metric. If the network has some SC core below the threshold, then defragmentation will occur at that point, which is applied only to cores with SC below the threshold. Defragmentation is done by repositioning a particular circuit in a different core, keeping the same slots index, or in the same core, changing the slots index. Defragmentation solutions are also proposed in \ac{SDM-EON} for scenarios with time multiplexing \cite{Zhao:19}.

In \cite{Yongli:16}, a technique called \textit{virtual concatenation} is proposed. With this model, the authors propose the allocation of slots of the same circuit in different cores, and in non-adjacent slots intervals, which makes it possible to mitigate the problem of fragmentation in this scenario. This approach is less discussed in the literature, since equipment has not yet been developed to support this type of allocation.

In order to evaluate some RMSCA algorithms found in literature, simulations were performed with the ONS simulator \cite{ONS-SBRC}. The independent replication method was employed to generate confidence intervals with 95\% confidence level. Each simulation run involved 100.000 requests with the following connection requests rates: 10, 20, 40, 80, 160 e 200 Gbps, all with the same arrival probability. Five load points were evaluated, and 5 replications were performed for each loading point. Connection requests follow a Poisson process with the mean holding time of 600 seconds, according to a negative exponential distribution and uniformly-distributed among all nodes-pairs.

\begin{figure}[!ht]
	\centering
	\includegraphics[width=0.9\textwidth]{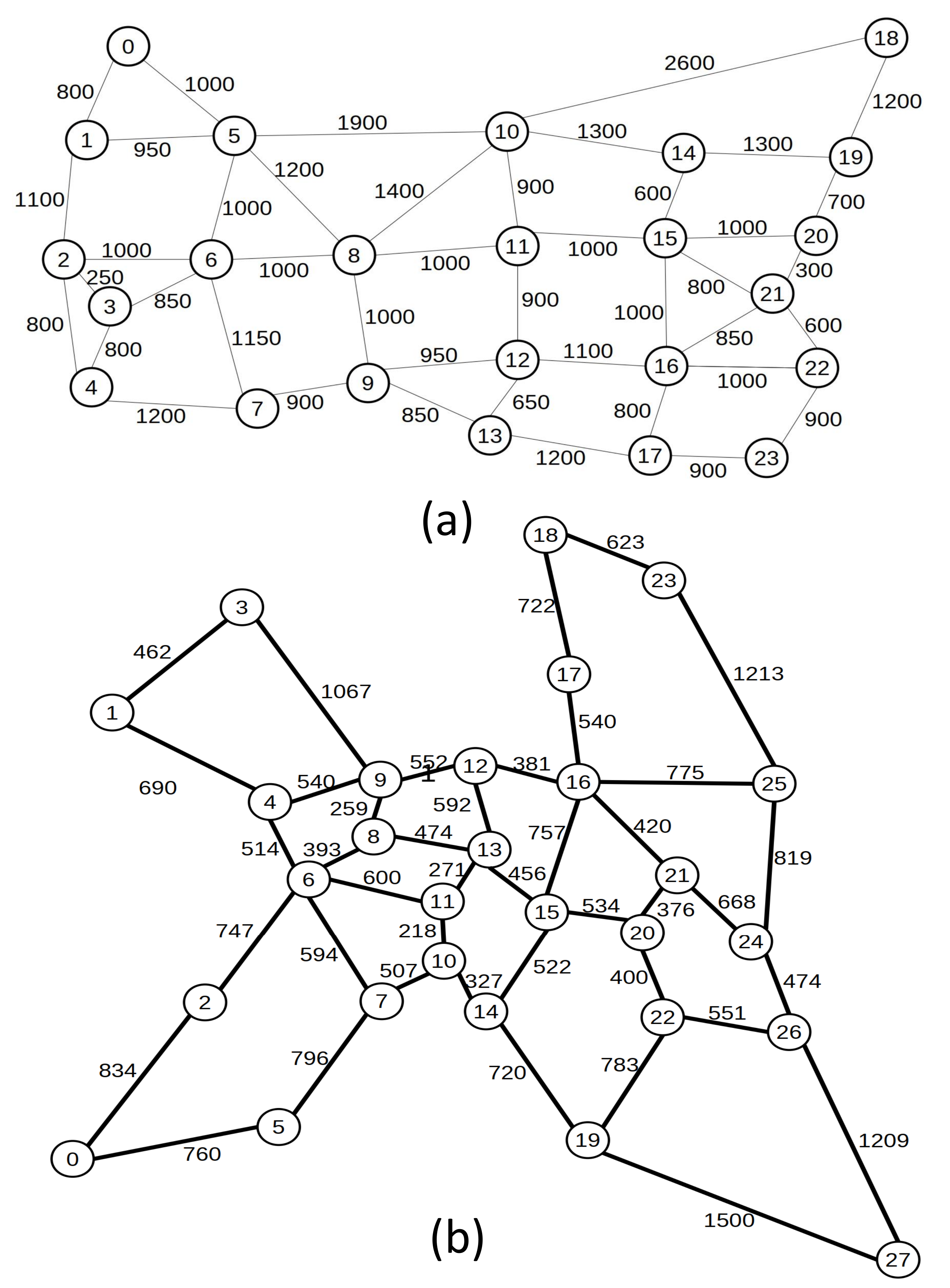}
	\caption{NSFNet (a) and USA (b) topologies.}
	\label{fig:topologias}
\end{figure}

The american USA topology (24 nodes and 43 bidirectional links) and european Paneuro (28 nodes and 41 bidirectional topology are used, shown in Figure \ref{fig:topologias}. The granularity of frequency slot is 12.5 GHz. Each fiber is a 7-MCF (fig. \ref{fig:mcf}(a)), with 320 slots in each core. The guard band between two adjacent lightpaths is assumed to be 1 slot.

Three algorithms were chosen for comparison. The first one, called \textit{Baseline}, is a classic model of the literature, characterized by the application of the Dijkstra (DJK) algorithm \cite{dijkstra:1959}  for routing and the \textit{FirstFit} allocation policy for slot and core selection. This algorithm is also used in some of the papers cited above \cite{Zhang:16}, \cite{Ruijie:16}. The second algorithm is RFCA \cite{Ruijie:16-2}, which consists of the random selection of spectral resources (cores and slots), always respecting the spectral continuity and contiguity constraints. The third algorithm is the Intra Area \cite{Tode:16}, in which the allocation of resources is performed taking into account the allocation in priority indexes of slots. The main idea is to show the comparison between algorithms that cause greater spectral fragmentation (RFCA), with an algorithm that prioritizes greater organization of the allocated resources (IntraArea), besides showing the performance of the two techniques compared with a classic algorithm of the literature (Baseline) that maintains a certain level of organization of resources (through allocation of resources by \ textit {FirstFit}).

The simulations are performed in a crosstalk-aware scenario, and different signal modulations are used. Table \ref{tab:modulation} present the available modulation formats, with the respective transmission rate and crosstalk threshold \cite{Ajmal:15}.

\begin{table}[]
	\caption{Modulation formats with the respective transmission rate per slot and crosstalk threshold.}
	\centering
	\label{tab:modulation}
	\begin{tabular}{|c|c|c|}
		\hline
		\rowcolor[HTML]{C0C0C0} 
		{\color[HTML]{333333} \textbf{Modulation}} & {\color[HTML]{333333} \textbf{\begin{tabular}[c]{@{}c@{}}Transmission\\ capacity\end{tabular}}} & {\color[HTML]{333333} \textbf{XT Threshold}} \\ \hline
		BPSK                                       & 12.5 Gbps                                                                                       & -14 dB                                       \\ \hline
		QPSK                                       & 25 Gbps                                                                                         & -18.5 dB                                     \\ \hline
		8QAM                                       & 37.5 Gbps                                                                                       & -21 dB                                       \\ \hline
		16QAM                                      & 50 Gbps                                                                                         & -25 dB                                       \\ \hline
		32QAM                                      & 62.5 Gbps                                                                                       & -27 dB                                       \\ \hline
		64QAM                                      & 75 Gbps                                                                                         & -34 dB                                       \\ \hline
	\end{tabular}
\end{table}

\begin{figure}[!ht]
	\centering
	\includegraphics[width=0.9\textwidth]{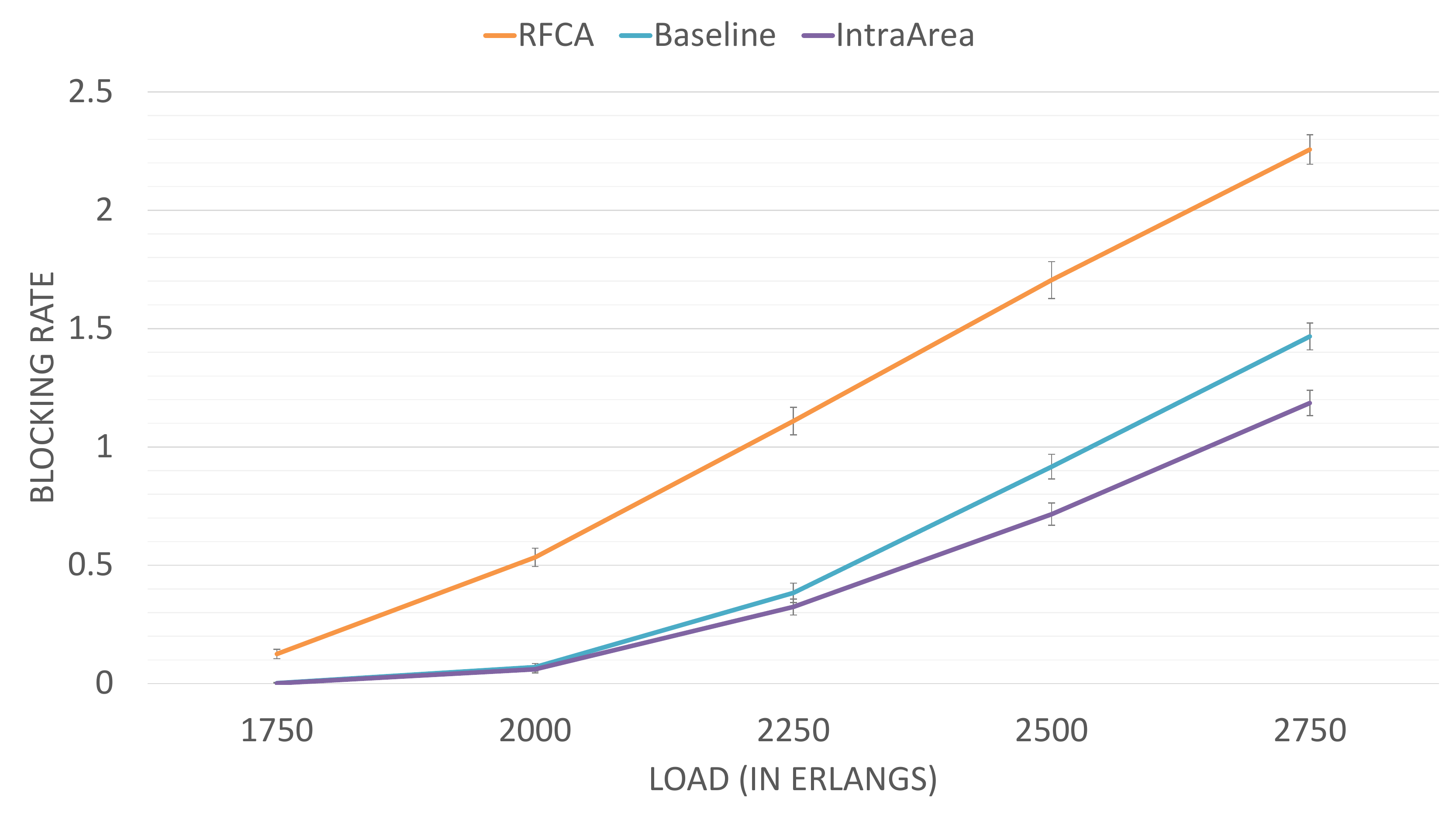}
	\caption{Blocking rate on USA topology.}
	\label{fig:brUSA}
\end{figure}

To perform the crosstalk evaluation, the following values are used in Equation \ref{eq:crosstalk}: $k=3.16 * 10^{-5}$, $r=55mm$, $\beta = 4 * 10^{6}$ e $w_{tr} = 45 \mu m$. The metrics evaluated were circuit block probability and bandwidth blocking probability. The figure \ref{fig:brUSA} shows the blocking rate for the USA topology.

The circuit blocking probability measures the number of circuits that have been blocked in the network, in relation to the total of circuits generated. It is observed that the Intra Area presents better performance in the evaluated scenario. It is demonstrated that the better organization of the allocated spectral resources makes possible the establishment of more circuit requests. The RFCA algorithm, by choosing randomly the cores and slots to be allocated, ends up causing more spectral fragmentation and consequently reaching a higher blocking rate, which is $ 53.86 \% $ higher than the closest performance competitor.

\begin{figure}[!ht]
	\centering
	\includegraphics[width=0.9\textwidth]{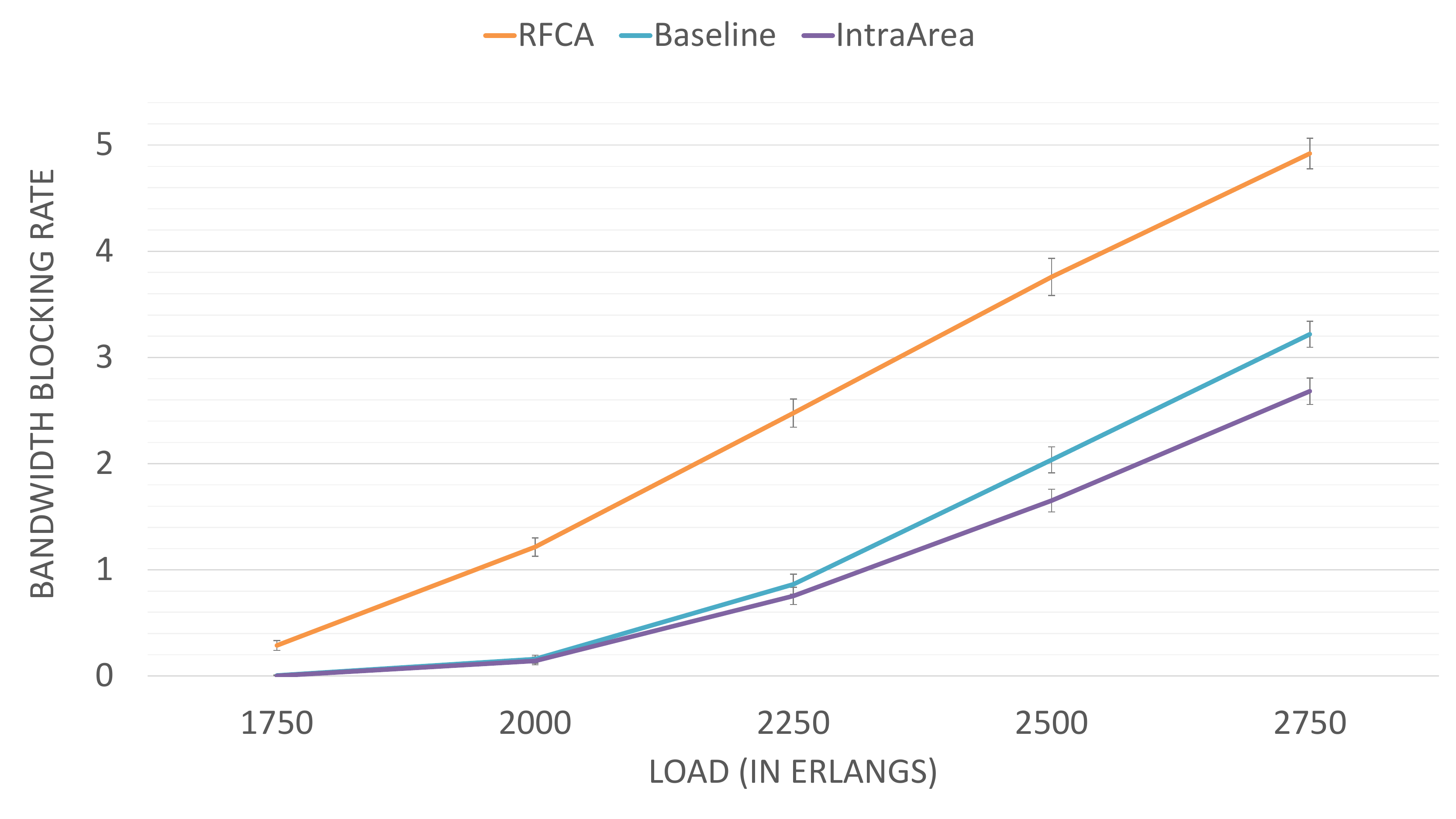}
	\caption{Bandwidth blocking rate on USA topology.}
	\label{fig:bbrUSA}
\end{figure}

The Intra Area and Baseline algorithms present similar performance in a low load scenario. The difference in performance increases with increasing load, and this is due to the reduced fragmentation caused by the Intra Area algorithm, since the free slot intervals within the priority areas have the number of slots suitable for circuits of the priority area in question, which facilitates the creation of new circuits. At the highest load point evaluated, there is a $ 20.18 \% $ reduction in the blocking caused by IntraArea when compared to the blocking caused by the Baseline. Figure \ref{fig:bbrUSA} shows the results of bandwidth blocking rate on USA topology.

The bandwidth blocking probability measures the total bandwidth blocked on the network, relative to the total bandwidth generated. When looking at the Figure \ref{fig:brUSA} and \ref{fig:bbrUSA} graphs, there is a small variation in the behavior of the graph. However, the blocking interval is higher for the bandwidth evaluation, since it is more common to block circuits with higher bandwidth demand than lower bandwidth circuits, which causes a greater impact on the bandwidth blocking rate. In relation to Intra Area, there is a reduction of $ 16.65 \% $ of the block when compared to the bandwidth blocking of the Baseline, and $ 45.49 \% $ when compared to the blocking of the RFCA.

\begin{figure}[!ht]
	\centering
	\includegraphics[width=0.9\textwidth]{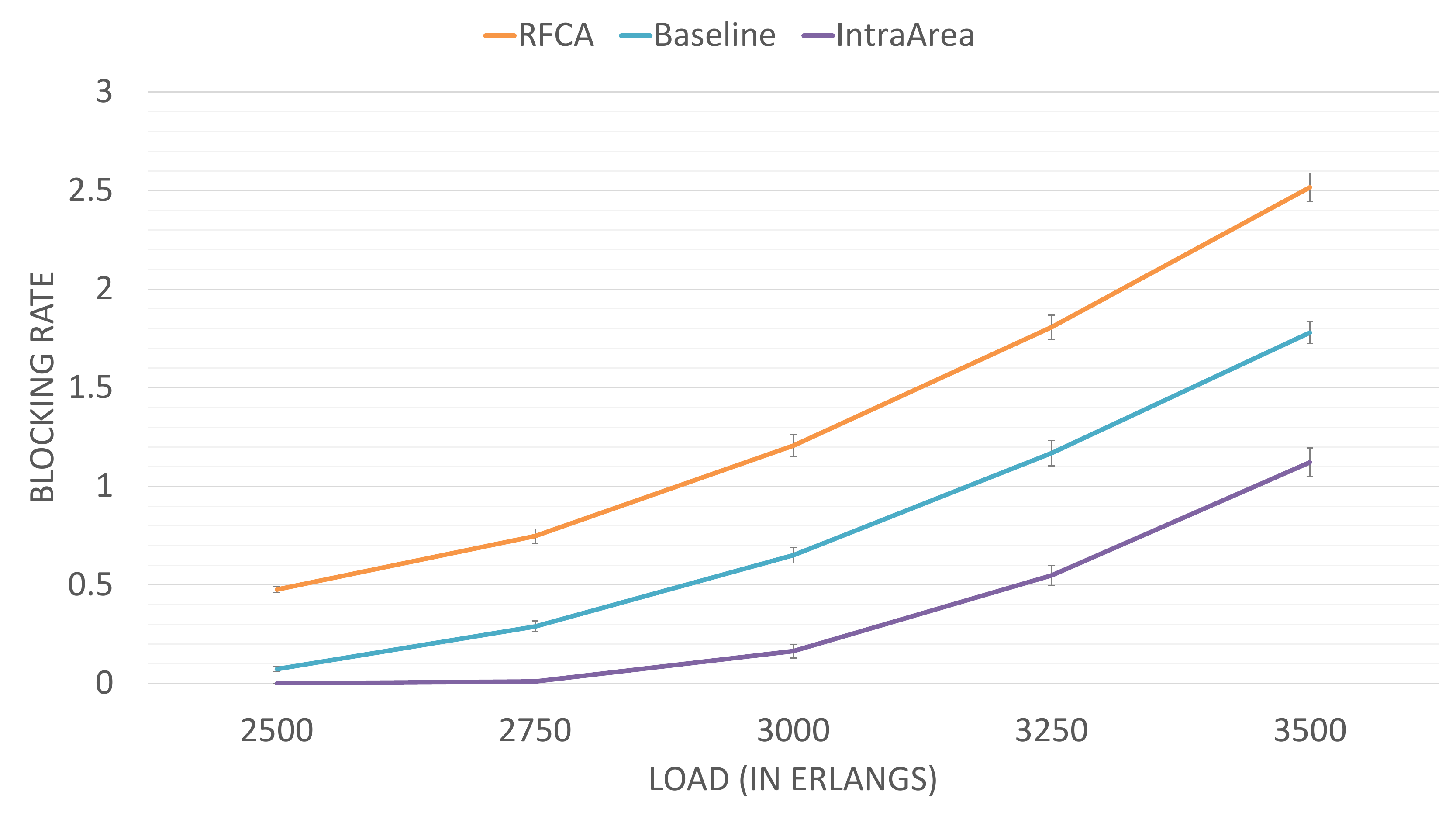}
	\caption{Blocking rate on Paneuro topology.}
	\label{fig:brPaneuro}
\end{figure}

Still, the spectral organization provided by the Intra Area algorithm allows a larger number of circuits to be accommodated in the spectrum, resulting in a lower blocking rate. The Baseline, in turn, forces the allocation of all circuits in the first cores, without differentiation as to the modulation adopted or number of slots of the circuits. Thus, when trying to allocate the central core, the new circuit suffers great crosstalk interference, since all periferic cores will already be allocated. Thus, in scenarios with crosstalk interference, the central core of the fibers has low spectral use in Baseline and RFCA, causing higher blocking rate. Besides that, the Intra Area has slot ranges reserved for each slot ratio of the circuits. Thus, it is easier to attend circuits of greater demand for bandwidth, as there will be spectral bands dedicated to them. Figure \ref{fig:brPaneuro} shows the results of blocking rate on Paneuro topology.

\begin{figure}[!ht]
	\centering
	\includegraphics[width=0.9\textwidth]{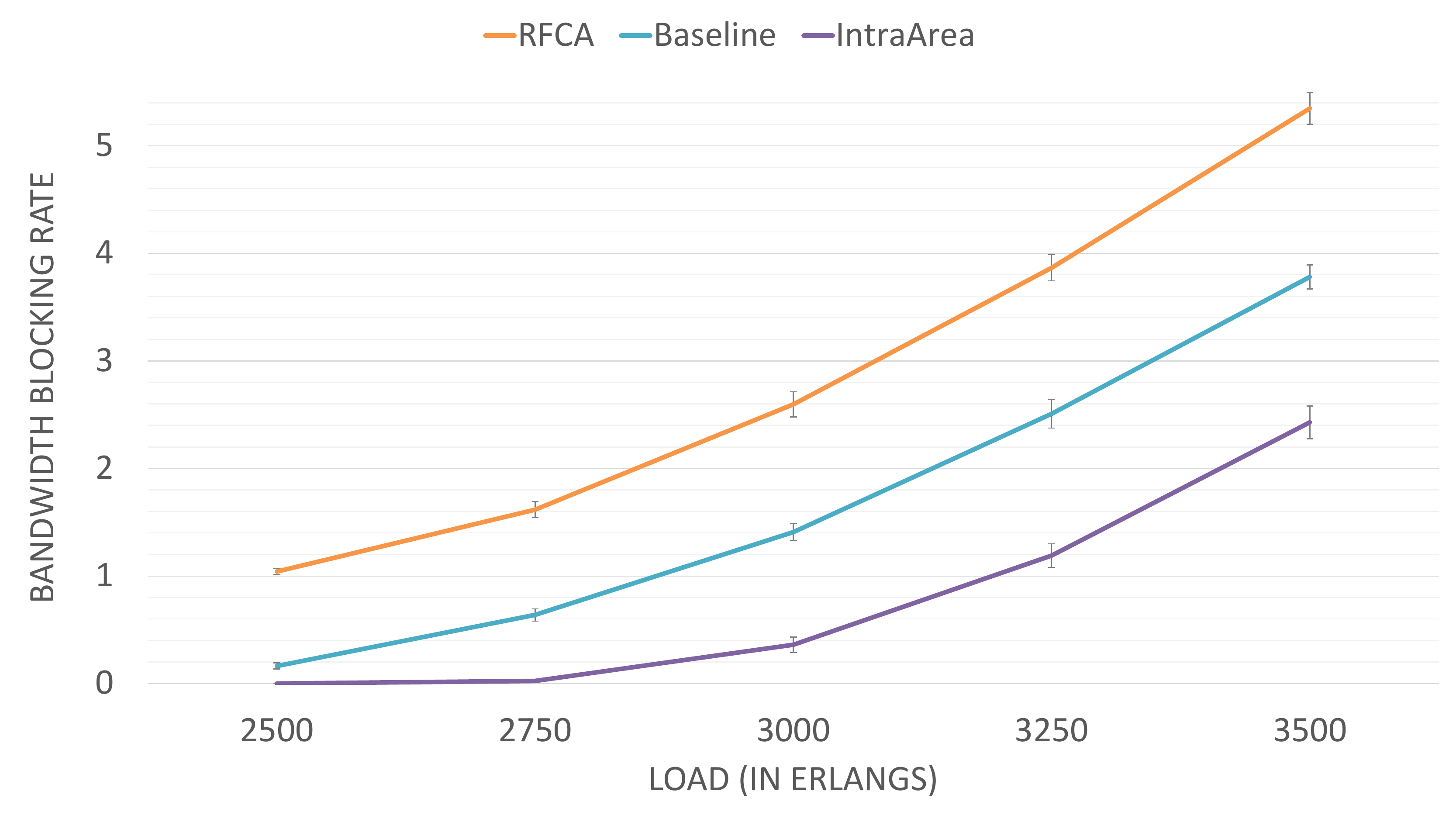}
	\caption{Bandwidth blocking rate on Paneuro topology.}
	\label{fig:bbrPaneuro}
\end{figure}

In the Paneuro topology, the Intra Area algorithm already achieves better performance from the lowest load point. Because it is a larger topology, there are more resources available (more links), requiring a higher load to observe the occurrence of blocking. The Intra Area algorithm continues to perform better, achieving a reduction of $ 36.97 \% $ on blocking rate when compared to the Baseline and $ 55.43 \% $ compared to the RFCA. The figure \ref{fig:bbrPaneuro} displays the bandwidth blocking rate result for the Paneuro topology.

A similar behavior is also observed between the graphs of Figures \ref{fig:brPaneuro} and \ref{fig:bbrPaneuro}. For the bandwidth blocking rate, the Intra Area algorithm presents blocking reduction of $ 35.79 \% $ compared to the Baseline and $ 54.6 \% $ compared to the RFCA.

The Intra Area algorithm achieves the best performance (lower blocking probability) while the RFCA achieves the worst performance in the evaluated scenarios. This is because the Intra Area keeps the spectrum more organized, making allocation attempts in the priorized and common areas. On the other hand, RFCA randomly allocates cores and slots through \textit{Random Fit} policy, causing fragmentation in the spectrum, and making it difficult to establish new circuits.

The next section presents the conclusions, challenges and open questions around \ac{SDM-EON} scenarios.

\section{Conclusions, challenges and prospects}\label{sec:conclusao}

The studies around \ac{SDM-EON} have grown in the literature, since \ac{MCF} appear as promising proposal to support the growing data traffic in \ac{EON}. The use of multi-core fibers ensures greater availability of resources as each core is compared to a single-core fiber of the traditional EON. However, the physical proximity of the cores in the fiber causes \textit{crosstalk} more accentuated in the \ac{SDM-EON} scenarios.

Considering the papers found so far in the literature of \ac{SDM-EON}, the data found were:
\begin{itemize}
	\item 15 papers propose solutions to solve the \ac{RMSCA} problem and the core allocation. Comparisons are made with classical algorithms of the EON literature, such as Dijkstra, KSP, First Fit and Random Fit;
	\item 17 papers use 7-core fibers. There are also studies with 3, 5, 8, 9, 12 and 19-core fibers;
	\item Although most papers (17 articles found) consider the \textit{crosstalk} effect on the link, only 2 portray other interferences of the physical medium;
	\item 16 papers use a dynamic traffic scenario, and 5 present a static traffic scenario;
	\item 4 papers consider protection and survival techniques in \ac{SDM-EON};
	\item 11 papers present the request generation with bandwidth demand between 1 and 10 slots, and 7 papers do not present traffic information.
	\item Only one paper does optical aggregation;
\end{itemize}

Thus, we can conclude that the most of the papers uses dynamic traffic configuration, which depicts a scenario closer to reality, since the circuit requests have source, destination and unknown traffic demand before the circuit establishment instant. In addition, there are many spectrum and core allocation solution proposals. However, these solutions are compared to classic literature algorithms, that are non-aware of crosstalk. A comparison between the new proposals has not yet been made. We can also note that the \textit{crosstalk} is a major problem in the \ac{SDM-EON} scenario, a concern found in most studies (17 papers). However, most (19 articles) ignore other interferences of the physical medium of propagation. Using information from the physical layer approximates the model performance to the real scenario of interference in data traffic. Few papers have been conducted in the area of survivability (4 articles) and traffic aggregation (1 article).

A major current challenge is the development of suitable equipment for switching the optical circuit in the SDM-EON. The papers found in the literature refer to a hypothetical architecture, capable of switching the signal between different cores (in some cases). The papers that present some evaluation about viable architectures \cite{Richardson:13}, \cite{Marom:17} cite as solution the adaptation of devices (switches, amplifiers and multiplexers) found in other types of networks.

Another challenge found in \ac{SDM-EON} is the mitigation of the \textit{crosstalk} effect. This interference of the physical layer is responsible for the unavailability of spectral resources, which are idle when crosstalk is high enough (\ac{XT} above $-25dB$ \cite{Richardson:13}) to make it impossible to allocate the circuit. Some types of fibers are proposed to reduce the effect of crosstalk (such as Trench-Assisted MCF), but there is still \textit{crosstalk} occurrence from a distance threshold defined by manufacturer.

It is also important to highlight as a challenge the production of equipment with a low financial cost, which allows the least expense for its implementation. In addition, the performance achieved by using a $n$ core \ac{MCF} should be similar to the performance obtained by $n$ coupled single-core fibers, which reinforces the development of fibers with greater tolerance to crosstalk interference.

When evaluating a scenario of \ac{SDM-EON} with occurrence of crosstalk, it is necessary to define accurately the characteristics related to the evaluation form of crosstalk. It has been demonstrated that a value of $ n $ can be used statically or dynamically, taking into account the crosstalk impact of the new circuit in circuits already established in the network. This variety of scenarios provides great impact on the network blocking rate, resulting in blocking differences of up to $ 74.13 \% $. A more in-depth study of the impact of different crosstalk scenarios can also be done in the future, in order to delineate the scenario closest to the occurrence of crosstalk in a real network.

The elaboration of a high efficiency (RMSCA) solution is also necessary. Nowadays, the papers found in the literature that propose RMSCA solutions compare their performance with classical literature algorithms such as the Dijkstra algorithm for routing and the First Fit strategy for spectrum allocation. No comparisons were made between the main proposed RMSCA solutions.

From our studies, it can be concluded that some scenarios are predominant in the literature, such as dynamic configuration of request generation, 7-core fibers, consideration of \textit{crosstalk}, and request bandwidth varying between 1 and 10 slots. It is also possible to observe many improvement opportunities, which will be explored as future works, such as the comparison between already proposed slot and core allocation techniques, the application of other physical layer effects besides \textit{crosstalk}, and the proposition of an impairment-aware allocation algorithm.

\bibliographystyle{elsarticle-num}
\bibliography{Surveys_Tutorials}

\begin{thebibliography}{10}
\expandafter\ifx\csname url\endcsname\relax
  \def\url#1{\texttt{#1}}\fi
\expandafter\ifx\csname urlprefix\endcsname\relax\def\urlprefix{URL }\fi
\expandafter\ifx\csname href\endcsname\relax
  \def\href#1#2{#2} \def\path#1{#1}\fi

\bibitem{Jinno:09}
M.~Jinno, H.~Takara, B.~Kozicki, Y.~Tsukishima, Y.~Sone, S.~Matsuoka,
  Spectrum-efficient and scalable elastic optical path network: architecture,
  benefits, and enabling technologies, IEEE Communications Magazine 47~(11)
  (2009) 66--73.
\newblock \href {http://dx.doi.org/10.1109/MCOM.2009.5307468}
  {\path{doi:10.1109/MCOM.2009.5307468}}.

\bibitem{Tode:14}
H.~Tode, Y.~Hirota, Routing, spectrum and core assignment for space division
  multiplexing elastic optical networks, in: 2014 16th International
  Telecommunications Network Strategy and Planning Symposium (Networks), 2014,
  pp. 1--7.
\newblock \href {http://dx.doi.org/10.1109/NETWKS.2014.6958538}
  {\path{doi:10.1109/NETWKS.2014.6958538}}.

\bibitem{Ajmal:15}
A.~Muhammad, G.~Zervas, R.~Forchheimer, Resource allocation for space-division
  multiplexing: Optical white box versus optical black box networking, Journal
  of Lightwave Technology 33~(23) (2015) 4928--4941.
\newblock \href {http://dx.doi.org/10.1109/JLT.2015.2493123}
  {\path{doi:10.1109/JLT.2015.2493123}}.

\bibitem{Drummond:17}
L.~R. Costa, A.~C. Drummond, New distance-adaptive modulation scheme for
  elastic optical networks, IEEE Communications Letters 21~(2) (2017) 282--285.
\newblock \href {http://dx.doi.org/10.1109/LCOMM.2016.2624288}
  {\path{doi:10.1109/LCOMM.2016.2624288}}.

\bibitem{Beyranvand:13}
H.~Beyranvand, J.~A. Salehi, A quality-of-transmission aware dynamic routing
  and spectrum assignment scheme for future elastic optical networks, Journal
  of Lightwave Technology 31~(18) (2013) 3043--3054.

\bibitem{Richardson:13}
D.~Richardson, J.~Fini, L.~Nelson, Space-division multiplexing in optical
  fibres, Nature Photonics 7~(5) (2013) 354--362.

\bibitem{Rademacher:17}
G.~Rademacher, R.~S. Lu\'{i}s, B.~J. Puttnam, Y.~Awaji, N.~Wada,
  \href{http://www.opticsexpress.org/abstract.cfm?URI=oe-25-10-12020}{Crosstalk
  dynamics in multi-core fibers}, Opt. Express 25~(10) (2017) 12020--12028.
\newblock \href {http://dx.doi.org/10.1364/OE.25.012020}
  {\path{doi:10.1364/OE.25.012020}}.
\newline\urlprefix\url{http://www.opticsexpress.org/abstract.cfm?URI=oe-25-10-12020}

\bibitem{Fini:11}
J.~M. Fini, B.~Zhu, T.~F. Taunay, M.~F. Yan, K.~S. Abedin, Crosstalk in
  multi-core optical fibres, in: 2011 37th European Conference and Exhibition
  on Optical Communication, 2011, pp. 1--3.

\bibitem{Amaya:13}
N.~Amaya, M.~Irfan, G.~Zervas, R.~Nejabati, D.~Simeonidou, J.~Sakaguchi,
  W.~Klaus, B.~Puttnam, T.~Miyazawa, Y.~Awaji, N.~Wada, I.~Henning,
  \href{http://www.opticsexpress.org/abstract.cfm?URI=oe-21-7-8865}{Fully-elastic
  multi-granular network with space/frequency/time switching using multi-core
  fibres and programmable optical nodes}, Opt. Express 21~(7) (2013)
  8865--8872.
\newblock \href {http://dx.doi.org/10.1364/OE.21.008865}
  {\path{doi:10.1364/OE.21.008865}}.
\newline\urlprefix\url{http://www.opticsexpress.org/abstract.cfm?URI=oe-21-7-8865}

\bibitem{Pouria:16}
P.~S. Khodashenas, J.~M. Rivas-Moscoso, D.~Siracusa, F.~Pederzolli,
  B.~Shariati, D.~Klonidis, E.~Salvadori, I.~Tomkos, Comparison of spectral and
  spatial super-channel allocation schemes for sdm networks, Journal of
  Lightwave Technology 34~(11) (2016) 2710--2716.
\newblock \href {http://dx.doi.org/10.1109/JLT.2016.2551299}
  {\path{doi:10.1109/JLT.2016.2551299}}.

\bibitem{Rui:16}
R.~Tian, Y.~Zhao, J.~Zhang, X.~Yu, Y.~Li, C.~Yu, J.~Zhang, C.~Liu, G.~Zhang,
  Dynamic traffic grooming based on auxiliary graph in spatial division
  multiplexing enabled elastic optical networks, in: 2016 15th International
  Conference on Optical Communications and Networks (ICOCN), 2016, pp. 1--3.
\newblock \href {http://dx.doi.org/10.1109/ICOCN.2016.7875878}
  {\path{doi:10.1109/ICOCN.2016.7875878}}.

\bibitem{Takenaga:11-2}
K.~Takenaga, Y.~Arakawa, Y.~Sasaki, S.~Tanigawa, S.~Matsuo, K.~Saitoh,
  M.~Koshiba,
  \href{http://www.opticsexpress.org/abstract.cfm?URI=oe-19-26-B543}{A large
  effective area multi-core fiber with an optimized cladding thickness}, Opt.
  Express 19~(26) (2011) B543--B550.
\newblock \href {http://dx.doi.org/10.1364/OE.19.00B543}
  {\path{doi:10.1364/OE.19.00B543}}.
\newline\urlprefix\url{http://www.opticsexpress.org/abstract.cfm?URI=oe-19-26-B543}

\bibitem{Takenaga:11}
K.~Takenaga, Y.~Arakawa, S.~Tanigawa, N.~Guan, S.~Matsuo, K.~Saitoh,
  M.~Koshiba, Reduction of crosstalk by trench-assisted multi-core fiber, in:
  2011 Optical Fiber Communication Conference and Exposition and the National
  Fiber Optic Engineers Conference, 2011, pp. 1--3.

\bibitem{Fujii:2014}
S.~Fujii, Y.~Hirota, H.~Tode, K.~Murakami, On-demand spectrum and core
  allocation for reducing crosstalk in multicore fibers in elastic optical
  networks, IEEE/OSA Journal of Optical Communications and Networking 6~(12)
  (2014) 1059--1071.
\newblock \href {http://dx.doi.org/10.1109/JOCN.2014.6985898}
  {\path{doi:10.1109/JOCN.2014.6985898}}.

\bibitem{savva:18}
G.~{Savva}, G.~{Ellinas}, B.~{Shariati}, I.~{Tomkos}, Physical layer-aware
  routing, spectrum, and core allocation in spectrally-spatially flexible
  optical networks with multicore fibers, in: 2018 IEEE International
  Conference on Communications (ICC), 2018, pp. 1--6.
\newblock \href {http://dx.doi.org/10.1109/ICC.2018.8422782}
  {\path{doi:10.1109/ICC.2018.8422782}}.

\bibitem{Jay:10}
J.~A. Jay, An overview of macrobending and microbending of optical fibers,
  White paper of Corning (2010) 1--21.

\bibitem{Hayashi:12}
T.~Hayashi, T.~Taru, O.~Shimakawa, T.~Sasaki, E.~Sasaoka, Characterization of
  crosstalk in ultra-low-crosstalk multi-core fiber, Journal of Lightwave
  Technology 30~(4) (2012) 583--589.
\newblock \href {http://dx.doi.org/10.1109/JLT.2011.2177810}
  {\path{doi:10.1109/JLT.2011.2177810}}.

\bibitem{Hayashi:12-2}
T.~Hayashi, T.~Taru, O.~Shimakawa, T.~Sasaki, E.~Sasaoka,
  \href{http://www.opticsexpress.org/abstract.cfm?URI=oe-20-26-B94}{Uncoupled
  multi-core fiber enhancing signal-to-noise ratio}, Opt. Express 20~(26)
  (2012) B94--B103.
\newblock \href {http://dx.doi.org/10.1364/OE.20.000B94}
  {\path{doi:10.1364/OE.20.000B94}}.
\newline\urlprefix\url{http://www.opticsexpress.org/abstract.cfm?URI=oe-20-26-B94}

\bibitem{Marom:17}
D.~M. Marom, P.~D. Colbourne, A.~D$'$Errico, N.~K. Fontaine, Y.~Ikuma,
  R.~Proietti, L.~Zong, J.~M. Rivas-Moscoso, I.~Tomkos, Survey of photonic
  switching architectures and technologies in support of spatially and
  spectrally flexible optical networking [invited], IEEE/OSA Journal of Optical
  Communications and Networking 9~(1) (2017) 1--26.
\newblock \href {http://dx.doi.org/10.1364/JOCN.9.000001}
  {\path{doi:10.1364/JOCN.9.000001}}.

\bibitem{Fontaine:13}
N.~K. Fontaine, Photonic lantern spatial multiplexers in space-division
  multiplexing, in: 2013 IEEE Photonics Society Summer Topical Meeting Series,
  2013, pp. 97--98.
\newblock \href {http://dx.doi.org/10.1109/PHOSST.2013.6614504}
  {\path{doi:10.1109/PHOSST.2013.6614504}}.

\bibitem{Ruijie:16-2}
R.~Zhu, Y.~Zhao, H.~Yang, H.~Chen, J.~Zhang, J.~P. Jue,
  \href{http://col.osa.org/abstract.cfm?URI=col-14-10-100604}{Crosstalk-aware
  rcsa for spatial division multiplexing enabled elastic optical networks with
  multi-core fibers}, Chin. Opt. Lett. 14~(10) (2016) 100604.
\newline\urlprefix\url{http://col.osa.org/abstract.cfm?URI=col-14-10-100604}

\bibitem{Yuanlong:16}
Y.~Tan, R.~Zhu, H.~Yang, Y.~Zhao, J.~Zhang, Z.~Liu, Q.~Qu, Z.~Zhou,
  Crosstalk-aware provisioning strategy with dedicated path protection for
  elastic multi-core fiber networks, in: 2016 15th International Conference on
  Optical Communications and Networks (ICOCN), 2016, pp. 1--3.
\newblock \href {http://dx.doi.org/10.1109/ICOCN.2016.7875849}
  {\path{doi:10.1109/ICOCN.2016.7875849}}.

\bibitem{Roberto:15}
R.~Proietti, L.~Liu, R.~P. Scott, B.~Guan, C.~Qin, T.~Su, F.~Giannone, S.~J.~B.
  Yoo, 3d elastic optical networking in the temporal, spectral, and spatial
  domains, IEEE Communications Magazine 53~(2) (2015) 79--87.
\newblock \href {http://dx.doi.org/10.1109/MCOM.2015.7045394}
  {\path{doi:10.1109/MCOM.2015.7045394}}.

\bibitem{Politi:12}
C.~T. Politi, V.~Anagnostopoulos, C.~Matrakidis, A.~Stavdas, A.~Lord,
  V.~López, J.~Fernández-Palacios, Dynamic operation of flexi-grid ofdm-based
  networks\href {http://dx.doi.org/10.1364/OFC.2012.OTh3B.2}
  {\path{doi:10.1364/OFC.2012.OTh3B.2}}.

\bibitem{ONS-SBRC}
L.~R. Costa, L.~S. de~Sousa, F.~R. de~Oliveira, K.~A. da~Silva, P.~J.~S.
  J\'unior, A.~C. Drummond,
  \href{http://sbrc2016.ufba.br/downloads/Salao_Ferramentas/154765.pdf}{{O}{N}{S}:
  {S}imulador de {E}ventos {D}iscretos para {R}edes \'{O}pticas
  {W}{D}{M}/{E}{O}{N}}, in: SBRC 2016 - Salao de Ferramentas, 2016.
\newline\urlprefix\url{http://sbrc2016.ufba.br/downloads/Salao_Ferramentas/154765.pdf}

\bibitem{Tode:16}
H.~Tode, Y.~Hirota, Routing, spectrum, and core and/or mode assignment on
  space-division multiplexing optical networks [invited], IEEE/OSA Journal of
  Optical Communications and Networking 9~(1) (2017) A99--A113.
\newblock \href {http://dx.doi.org/10.1364/JOCN.9.000A99}
  {\path{doi:10.1364/JOCN.9.000A99}}.

\bibitem{Muhammad:14}
A.~Muhammad, G.~Zervas, D.~Simeonidou, R.~Forchheimer, Routing, spectrum and
  core allocation in flexgrid sdm networks with multi-core fibers, in: 2014
  International Conference on Optical Network Design and Modeling, 2014, pp.
  192--197.

\bibitem{Seitaro:17}
S.~Sugihara, Y.~Hirota, S.~Fujii, H.~Tode, T.~Watanabe, Dynamic resource
  allocation for immediate and advance reservation in
  space-division-multiplexing-based elastic optical networks, IEEE/OSA Journal
  of Optical Communications and Networking 9~(3) (2017) 183--197.
\newblock \href {http://dx.doi.org/10.1364/JOCN.9.000183}
  {\path{doi:10.1364/JOCN.9.000183}}.

\bibitem{Wang:14}
R.~Wang, B.~Mukherjee,
  \href{http://www.sciencedirect.com/science/article/pii/S1573427713000799}{Spectrum
  management in heterogeneous bandwidth optical networks}, Optical Switching
  and Networking 11~(Part A) (2014) 83 -- 91.
\newblock \href {http://dx.doi.org/https://doi.org/10.1016/j.osn.2013.09.003}
  {\path{doi:https://doi.org/10.1016/j.osn.2013.09.003}}.
\newline\urlprefix\url{http://www.sciencedirect.com/science/article/pii/S1573427713000799}

\bibitem{Zhang:16}
L.~Zhang, N.~Ansari, A.~Khreishah, Anycast planning in space division
  multiplexing elastic optical networks with multi-core fibers, IEEE
  Communications Letters 20~(10) (2016) 1983--1986.
\newblock \href {http://dx.doi.org/10.1109/LCOMM.2016.2593479}
  {\path{doi:10.1109/LCOMM.2016.2593479}}.

\bibitem{Tode:13}
S.~Fujii, Y.~Hirota, H.~Tode, Dynamic resource allocation with virtual grid for
  space division multiplexed elastic optical network, in: 39th European
  Conference and Exhibition on Optical Communication (ECOC 2013), 2013, pp.
  1--3.
\newblock \href {http://dx.doi.org/10.1049/cp.2013.1653}
  {\path{doi:10.1049/cp.2013.1653}}.

\bibitem{Oliveira:17}
H.~M.~N. da~Silva~Oliveira, N.~L.~S. da~Fonseca, The minimum interference
  p-cycle algorithm for protection of space division multiplexing elastic
  optical networks, IEEE Latin America Transactions 15~(7) (2017) 1342--1348.
\newblock \href {http://dx.doi.org/10.1109/TLA.2017.7959516}
  {\path{doi:10.1109/TLA.2017.7959516}}.

\bibitem{Hashino:17}
K.~Hashino, Y.~Hirota, Y.~Tanigawa, H.~Tode,
  \href{http://www.osapublishing.org/abstract.cfm?URI=PS-2017-PM3D.2}{Crosstalk-aware
  spectrum and core allocation with crosstalk-prohibited frequency slot in
  space-division multiplexing elastic optical networks}, in: Advanced Photonics
  2017 (IPR, NOMA, Sensors, Networks, SPPCom, PS), Optical Society of America,
  2017, p. PM3D.2.
\newblock \href {http://dx.doi.org/10.1364/PS.2017.PM3D.2}
  {\path{doi:10.1364/PS.2017.PM3D.2}}.
\newline\urlprefix\url{http://www.osapublishing.org/abstract.cfm?URI=PS-2017-PM3D.2}

\bibitem{Oliveira:16}
H.~M. N.~S. Oliveira, N.~L.~S. da~Fonseca, Algorithm for protection of space
  division multiplexing elastic optical networks, in: 2016 IEEE Global
  Communications Conference (GLOBECOM), 2016, pp. 1--6.
\newblock \href {http://dx.doi.org/10.1109/GLOCOM.2016.7841575}
  {\path{doi:10.1109/GLOCOM.2016.7841575}}.

\bibitem{Oliveira:17-2}
H.~M. N.~S. Oliveira, N.~L.~S. da~Fonseca, Algorithm for shared path for
  protection of space division multiplexing elastic optical networks, in: 2017
  IEEE International Conference on Communications (ICC), 2017, pp. 1--6.
\newblock \href {http://dx.doi.org/10.1109/ICC.2017.7997378}
  {\path{doi:10.1109/ICC.2017.7997378}}.

\bibitem{Ruijie:16}
R.~Zhu, Y.~Zhao, H.~Yang, X.~Yu, Y.~Tan, J.~Zhang, N.~Wang, J.~P. Jue,
  Multi-dimensional resource assignment in spatial division multiplexing
  enabled elastic optical networks with multi-core fibers, in: 2016 15th
  International Conference on Optical Communications and Networks (ICOCN),
  2016, pp. 1--3.
\newblock \href {http://dx.doi.org/10.1109/ICOCN.2016.7875672}
  {\path{doi:10.1109/ICOCN.2016.7875672}}.

\bibitem{Meloni:16}
G.~{Meloni}, F.~{Fresi}, M.~{Imran}, F.~{Paolucci}, F.~{Cugini},
  A.~{D’Errico}, L.~{Giorgi}, T.~{Sasaki}, P.~{Castoldi}, L.~{Pot},
  Software-defined defragmentation in space-division multiplexing with
  quasi-hitless fast core switching, Journal of Lightwave Technology 34~(8)
  (2016) 1956--1962.
\newblock \href {http://dx.doi.org/10.1109/JLT.2015.2503434}
  {\path{doi:10.1109/JLT.2015.2503434}}.

\bibitem{Imran:15}
M.~{Imran}, F.~{Paolucci}, F.~{Cugini}, A.~{D'Errico}, L.~{Giorgi},
  T.~{Sasaki}, P.~{Castoldi}, L.~{Poti}, Quasi-hitless software-defined
  defragmentation in space division multiplexing (sdm), in: 2015 European
  Conference on Optical Communication (ECOC), 2015, pp. 1--3.
\newblock \href {http://dx.doi.org/10.1109/ECOC.2015.7341673}
  {\path{doi:10.1109/ECOC.2015.7341673}}.

\bibitem{Zhao:18}
Y.~{Zhao}, L.~{Hu}, R.~{Zhu}, X.~{Yu}, X.~{Wang}, J.~{Zhang}, Crosstalk-aware
  spectrum defragmentation based on spectrum compactness in space division
  multiplexing enabled elastic optical networks with multicore fiber, IEEE
  Access 6 (2018) 15346--15355.
\newblock \href {http://dx.doi.org/10.1109/ACCESS.2018.2795102}
  {\path{doi:10.1109/ACCESS.2018.2795102}}.

\bibitem{Zhao:19}
Y.~Zhao, L.~Hu, R.~Zhu, X.~Yu, Y.~Li, W.~Wang, J.~Zhang,
  \href{http://www.opticsexpress.org/abstract.cfm?URI=oe-27-4-5014}{Crosstalk-aware
  spectrum defragmentation by re-provisioning advance reservation requests in
  space division multiplexing enabled elastic optical networks with multi-core
  fiber}, Opt. Express 27~(4) (2019) 5014--5032.
\newblock \href {http://dx.doi.org/10.1364/OE.27.005014}
  {\path{doi:10.1364/OE.27.005014}}.
\newline\urlprefix\url{http://www.opticsexpress.org/abstract.cfm?URI=oe-27-4-5014}

\bibitem{Yongli:16}
Y.~Zhao, J.~Zhang, Crosstalk-aware cross-core virtual concatenation in spatial
  division multiplexing elastic optical networks, Electronics Letters 52~(20)
  (2016) 1701--1703.
\newblock \href {http://dx.doi.org/10.1049/el.2016.2132}
  {\path{doi:10.1049/el.2016.2132}}.

\bibitem{dijkstra:1959}
E.~W. Dijkstra, A note on two problems in connexion with graphs, Numerische
  mathematik 1~(1) (1959) 269--271.

\end{thebibliography}

\end{document}